# Gradient waveform design for tensor-valued encoding in diffusion MRI

Filip Szczepankiewicz[1,2,3,*], Carl-Fredrik Westin[1,2] and Markus Nilsson[3]

1. Radiology, Brigham and Women's Hospital, Boston, MA, US
2. Harvard Medical School, Boston, MA, US
3. Clinical Sciences, Lund, Lund University, Lund, Sweden

* Corresponding author: FS (filip.szczepankiewicz@med.lu.se)

**Abstract -** Diffusion encoding along multiple spatial directions per signal acquisition can be described in terms of a b-tensor. The benefit of tensor-valued diffusion encoding is that it unlocks the 'shape of the b-tensor' as a new encoding dimension. By modulating the b-tensor shape, we can control the sensitivity to microscopic diffusion anisotropy which can be used as a contrast mechanism; a feature that is inaccessible by conventional diffusion encoding. Since imaging methods based on tensor-valued diffusion encoding are finding an increasing number of applications we are prompted to highlight the challenge of designing the optimal gradient waveforms for any given application. In this review, we first establish the basic design objectives in creating field gradient waveforms for tensor-valued diffusion MRI. We also survey additional design considerations related to limitations imposed by hardware and physiology, potential confounding effects that cannot be captured by the b-tensor, and artifacts related to the diffusion encoding waveform. Throughout, we discuss the expected compromises and tradeoffs with an aim to establish a more complete understanding of gradient waveform design and its impact on accurate measurements and interpretations of data.

## 1. Introduction and background

Diffusion magnetic resonance imaging (dMRI) sensitizes the MR signal to the random movement of MR-visible particles, most commonly the hydrogen nuclei in water molecules. As the water moves randomly throughout the tissue, it probes the local environment and senses hindrances and restrictions imposed by the tissue microstructure. In this process, certain features of the microstructure are impressed on the movement of the water and—given an appropriate experiment—can be inferred from the observed MR signal. For example, the apparent diffusivity is reduced as tissue density increases (Chen et al., 2013) and may be anisotropic if the structures allow faster diffusion along a given direction (Stanisz et al., 1997, Beaulieu, 2002). Thus, dMRI provides a non-invasive—albeit indirect—probe of the tissue microstructure (Novikov et al., 2019, Jelescu et al., 2020). A major discovery that propelled the use of dMRI as a clinical and research tool, was that it could detect cerebral ischemia at a considerably earlier stage than other contemporary imaging modalities (Moseley et al., 1990a, Moseley et al., 1990b). Since then, dMRI has been used in a wide range of clinical applications, such as diagnostics of stroke, detection and staging of tumors in the central nervous system, prostate and breast, as well as for planning surgery and radiotherapy (Sundgren et al., 2004, Tsien et al., 2014, Partridge et al., 2017). It has also been useful in medical research, creating a better understanding of brain development (Lebel et al., 2019), learning (Zatorre et al., 2012, Thomas and Baker, 2013), white matter morphology and connectivity (Jones, 2008, Tournier, 2019), development of cancers (Padhani et al., 2009, Nilsson et al., 2018) and diseases of the body (Horsfield and Jones, 2002, Jellison et al., 2004, Taouli et al., 2016, Budde and Skinner, 2018, Assaf et al., 2019).

The vast majority of past and present dMRI studies are based on the canonical design proposed by Stejskal and Tanner (1965), where a pair of trapezoidal pulsed field gradients flank the refocusing pulse in a spin-echo sequence (Hahn, 1950). We call this design single diffusion encoding (SDE), using the convention described by Shemesh et al. (2016). However, in its simplicity, it has a major drawback. Since it can only apply diffusion encoding along one direction per signal readout, it is intrinsically bound to conflate three distinct and ubiquitous features of diffusion, namely the microscopic diffusion anisotropy, orientation dispersion and heterogeneity of isotropic diffusivities (Mitra, 1995, Novikov et al., 2018c, Henriques et al., 2019), making it a poor probe of microscopic diffusion anisotropy in complex tissue (Szczepankiewicz et al., 2015). We emphasize that this limitation does not arise due to the choice of a particular analysis method but is rather a fundamental limitation of the information encoded by the measurement itself (Mitra, 1995). Thus, a more elaborate experimental design is warranted.

The first solution to the problem was introduced by Cory et al. (1990), who showed that double diffusion encoding (DDE) could measure local pore geometry even if the substrate was not orientation coherent on the scale of a voxel. As indicated by the name, DDE is an extension of SDE that introduces a second pair of diffusion encoding pulses. Although DDE had been used before, the novelty was that the encoding direction of each pair could be controlled independently so that diffusivity along two directions could be probed in a single preparation of the signal. In essence, microscopic diffusion anisotropy creates a relation between encoding directions and the measured MR signal. When DDE with parallel and orthogonal pulse directions are compared, a higher diffusion anisotropy manifests as a larger difference in signal. The principles of DDE-based methods have been excellently described by Özarslan (2009), Shemesh et al. (2010b), Finsterbusch (2011), and Callaghan (2011). The conception of DDE spawned a rich field that still explores non-conventional diffusion encoding as a probe of features that are inaccessible by conventional means. Most notably, it has been used to disentangle microscopic diffusion anisotropy from orientation dispersion (Callaghan and Komlosh, 2002, Özarslan and Basser, 2008, Lawrenz et al., 2010, Jespersen et al., 2013, Jensen et al., 2014, Shemesh et al., 2010a, Komlosh et al., 2007, Najac et al., 2019, Ianus et al., 2016). The versatility of DDE has also made it a vital tool in the study of diffusion time (Clark et al., 2001, Ozarslan and Basser, 2007, Shemesh et al., 2009), exchange (Callaghan and Furó, 2004), motion (Ahn et al., 1987, Caprihan and Fukushima, 1990) and tractography (Wedeen et al., 2012), although these predominately use it for encoding along a single direction per signal acquisition.

A similar probe of microscopic diffusion anisotropy can also be achieved by combining conventional diffusion encoding along a single direction with isotropic diffusion encoding. This particular combination was proposed as a probe of microscopic diffusion anisotropy by Eriksson et al. (2013) and a method for quantification was developed by Lasič et al. (2014). In analogy to DDE-based methods, the divergence between signal acquired using the two types of diffusion encoding was related to the presence of microscopic diffusion anisotropy. Moreover, by investigating the non-monoexponential decay of the signal acquired with isotropic encoding, the heterogeneity of isotropic diffusivities could be estimated, i.e., the presence of multiple compartments with different bulk diffusivities (Lasič et al., 2014, Szczepankiewicz et al., 2015). However, the analysis lacked the ability to natively include the effects of macroscopic diffusion anisotropy and was not compatible with other gradient waveform designs, such as DDE.

A general framework for describing diffusion encoding for arbitrary gradient waveforms and their effect on multi-Gaussian diffusion was proposed by Westin et al. (2014), thereby bringing all waveform designs under one roof. In this framework, the conventional description of the experiment in terms of the b-value and encoding direction (Le Bihan et al., 1986) was replaced by the 'b-tensor' which added the *shape* of the diffusion encoding to the description (Westin et al., 2014, Eriksson et al., 2015, Westin et al., 2016). Although conventional diffusion encoding was already routinely described by a rank-1 'b-matrix,' even including the higher-rank contribution from imaging gradients (Mattiello et al., 1997), the novelty of this framework was to purposefully execute measurements with b-tensors of varying shape to control the measurement's sensitivity to diffusion anisotropy. For example, encoding with linear b-tensors (conventional) is maximally sensitive to diffusion anisotropy on both macro- and microscopic scales, spherical b-tensors are entirely insensitive to diffusion anisotropy, and intermediate b-tensor shapes have a sensitivity that depends on the configuration of their eigenvalues (Westin et al., 2016, Topgaard, 2017). This meant that SDE, DDE and arbitrary gradient modulation were all plausible candidates in terms of gradient waveform designs for b-tensors of rank up to 1, 2 and 3, respectively. Since the b-tensor framework requires diffusion encoding in more than one direction per signal preparation—a measurement that cannot be fully described by a b-value and encoding direction—we refer to such encoding as 'tensor-valued.'

The b-tensor formalism and tensor-valued diffusion encoding has seen a rapid uptake in dMRI research. Early applications include microstructure imaging of healthy brain (Szczepankiewicz et al., 2015, Dhital et al., 2018, Dhital et al., 2019, Lampinen et al., 2019, Tax et al., 2020), brain tumors (Szczepankiewicz et al., 2016, Nilsson et al., 2020a), multiple sclerosis (Winther Andersen et al., 2020), brain lesions related to epilepsy (Lampinen et al., 2020b), as well as body applications in kidney (Nery et al., 2019) and heart (Lasic et al., 2020). It has been demonstrated to improve quantification of fiber dispersion (Cottaar et al., 2020), biophysical compartment modelling (Lampinen et al., 2019, Afzali et al., 2019, Reisert et al., 2019), and it adds information to diffusion-relaxation-correlation experiments that improves the separability of water pools in heterogeneous biological tissue (Lampinen et al., 2020a, de Almeida Martins et al., 2020).

The feasibility of deploying b-tensor encoding on clinical systems hinges on efficient gradient waveform design that can deliver the required b-tensor in a short encoding time. In addition to DDE, gradient waveform designs that yield tensor-valued diffusion encoding have been proposed in related contexts. Most notably, isotropic diffusion encoding was introduced independently by Mori and van Zijl (1995) and Wong et al. (1995) for the purpose of rapid measurement of the mean diffusivity; as few as two images could be used to estimate the mean diffusivity in an anisotropic substrate instead of the minimal of four images when using SDE (Mori and van Zijl, 1995, Wong et al., 1995, Butts et al., 1997, Heid and Weber, 1997, Moffat et al., 2004, Ianus and Shemesh, 2017). More recently, the paradigm of using trapezoidal pulse shapes was replaced by freely modulated gradient waveforms which allows more freedom in the design (Drobnjak and Alexander, 2011, Truffet et al., 2020). For example, Topgaard (2013) introduced isotropic weighting by modulating the gradients so that the dephasing vector (q-vector) was spun at the magic angle, trading encoding efficiency for lower energy consumption and heating (Topgaard, 2016). A more general, and highly versatile, optimization framework was developed by Sjölund et al. (2015). This framework can optimize gradient waveforms for tensor-valued diffusion encoding tailored to any given pulse sequence timing, requirements of the hardware and subsequent analysis. This provides gradient waveforms with superior encoding efficiency—in some cases reducing the echo time by a factor of two—that facilitate minimal echo times and data quality that is comparable to routine diffusion MRI (Szczepankiewicz et al., 2019c).

Evidently, there exist many capable gradient waveform designs that are useful for tensor-valued diffusion encoding. However, selecting the optimal combination of waveforms remains a challenge; some designs are more useful than others, and future experiments may incorporate previously unexplored design goals. We therefore survey past and present designs that can be used for tensor-valued diffusion encoding and explore several aspects that may inform the design of ever more efficient and/or specialized gradient waveforms. As a touchstone, we show a selection of waveforms in Figure 1; all adapted to a common premise described in the caption. The figure shows waveforms that produce b-tensors with different shapes, encoding efficiency, trajectories through q-space, limits on the gradient vector magnitude, compensation for concomitant gradient effects and energy consumption—all of which will be described and referenced throughout this review. Since visualization of some of these features translates poorly onto paper, we have shared all waveforms and figures so that they can be manipulated and enjoyed in three-dimensions[1].

In this review, we will establish general goals for gradient waveform design and how they relate to limitations imposed by the MRI hardware as well as human physiology. We also provide an overview of the features that cannot be captured by the b-tensor, and may therefore be exploited in extended experiments, or, if ignored, impact the measurement as confounders. Lastly, we describe common artifacts that interact with the gradient waveform design, and we provide remedies where possible.

[1] Waveforms and figures are created/stored in MATLAB format (The MathWorks, Natick, MA, R2019a), and can be downloaded from https://github.com/filip-szczepankiewicz/Szczepankiewicz_JNeuMeth_2021.

## 2. Tensor-valued diffusion encoding

### *2.1. Theory of tensor-valued diffusion encoding*

Diffusion weighting is achieved by inducing phase incoherence in an ensemble of spins that exhibit random movement (Torrey, 1956, Pipe, 2010). Collections of spins can be represented by magnetization vectors, where the phase of each vector depends on the magnetic field strength experienced over a given time. By applying a magnetic field gradient over the object, we create a connection between movement and phase shift. The phase shift is proportional to the strength of the applied gradient and the distance that the spin traversed along the direction of the gradient, which can also be thought of as movement along a Larmor frequency gradient. For any given gradient waveform ($\mathbf{g}(t)$), the phase ($\phi$) at the echo time ($\tau$) can be expressed as

$$\phi = \gamma \int_0^\tau \mathbf{r} \cdot \mathbf{g}(t) \mathrm{d}t, \qquad \text{Eq. 1}$$

where $\mathbf{r}(t)$ is the position at time $t$, $\gamma$ is the gyromagnetic ratio (Price, 1997, Stejskal and Tanner, 1965) and '·' is the scalar dot product. The signal from any given ensemble of spins is simply the average of all spin vectors, which can be written as

$$S = \mathrm{S}_0 \langle \exp(-i\phi) \rangle, \qquad \text{Eq. 2}$$

where $\langle \cdot \rangle$ denotes averaging over the ensemble. From Eq. 2 we see that the signal is attenuated as the distribution of phases becomes more incoherent, a process that is expedited by faster incoherent motion and/or stronger gradients. We emphasize that phase coherency can be lost due to mechanisms other than diffusion, indeed, any incoherent motion will do (Le Bihan et al., 1986, Ahn et al., 1987, Butts et al., 1996). Bulk motion, on the other hand, shifts the global phase of the ensemble without reducing the signal magnitude (Hahn, 1960, Stejskal, 1965, Moran, 1982). For approximately Gaussian diffusion, Eq. 2 can be approximated by the cumulant expansion (Stepisnik, 1999, Grebenkov, 2007), such that the magnitude of the diffusion weighted signal will depend on the variance ($\langle \phi^2 \rangle$) of the phase distribution

$$S \approx \mathrm{S}_0 \exp\left(-\frac{\langle \phi^2 \rangle}{2}\right) = \mathrm{S}_0 \exp(-\mathbf{B}\colon \mathbf{D}), \qquad \text{Eq. 3}$$

where $\mathbf{B}$ is the diffusion encoding tensor (b-tensor or b-matrix) (Westin et al., 2014), ':' is the double inner product, and $\mathbf{D}$ is the diffusion tensor (Stejskal, 1965). It is worth noting, that diffusion encoding along a single direction (e.g. SDE) can be written in terms of a one-dimensional gradient waveform with encoding strength ($b$) and direction $\mathbf{n}$ (unit vector), such that $\mathbf{B} = b\ \mathbf{n}\otimes\mathbf{n}$ is a rank-1 tensor[2], where '$\otimes$' denotes the outer product. However, the use of diffusion encoding along multiple directions per shot does not conform to this description, prompting the b-tensor formalism which allows b-tensors of rank up to 3 (Westin et al., 2014), i.e., 'tensor-valued' diffusion encoding. The b-tensor is calculated from the effective gradient waveform[3], specified as a gradient trajectory defined on three orthogonal axes, such that

$$\mathbf{g}(t) = [g_\mathrm{x}(t) \quad g_\mathrm{y}(t) \quad g_\mathrm{z}(t)]^\mathrm{T}. \qquad \text{Eq. 4}$$

The b-tensor, or b-matrix (Mattiello et al., 1997), is the time integral over the outer product of the dephasing q-vector, such that (Westin et al., 2014, Westin et al., 2016)

$$\mathbf{B} = \int_0^\tau \mathbf{q}(t)\otimes\mathbf{q}(t)\mathrm{d}t, \qquad \text{Eq. 5}$$

where '$\otimes$' denotes the outer product, and the q-vector is

$$\mathbf{q}(t) = \gamma \int_0^t \mathbf{g}(t')\mathrm{d}t'. \qquad \text{Eq. 6}$$

For convenience, we may define rotation invariant metrics by which we describe the b-tensor. The conventional b-value, i.e., strength of the encoding, is its trace

$$b = \mathrm{Tr}(\mathbf{B}), \qquad \text{Eq. 7}$$

whereas the orientation (or direction) of $\mathbf{B}$ is captured by its eigenvectors, and its complete shape is defined by its eigenvalues ($\lambda_i$). If the b-tensor is axisymmetric[4] we can describe its shape by the b-tensor anisotropy ($b_\Delta$), defined as (Eriksson et al., 2015)

$$b_\Delta = (\lambda_\parallel - \lambda_\perp)/b, \qquad \text{Eq. 8}$$

where $\lambda_\parallel$ and $\lambda_\perp$ are the axial and radial eigenvalues. $b_\Delta$ is in the interval [–0.5 0) for oblate and (0 1] for prolate b-tensors, whereas $b_\Delta = 0$ for spherical b-tensors. Naturally, these metrics are analogous to rotation invariant metrics derived from the diffusion tensor (Kingsley, 2006, Basser et al., 1994, Westin et al., 2002).

---

[2] The use of 'rank' and 'order' is discrepant across literature. Here, we use rank to mean the number of dimensions spanned by the column vectors, whereas order is the number of indices necessary to address the elements. For example, b-tensors are of order 2 and have a rank between 0 and 3.

[3] Throughout this paper, we use/show 'effective gradient waveforms' which include the effects of refocusing pulses. Therefore, two monopolar pulses played along the same physical direction will be shown to have opposite directions if separated by a refocusing pulse.

[4] Axisymmetric b-tensors have at most two unique eigenvalues ($\lambda_\parallel$ and $\lambda_\perp$), such that one of the values is repeated twice ($\lambda_\perp$). The axis of symmetry is along the unique eigenvalue ($\lambda_\parallel$) which is the smallest/largest for oblate/prolate tensors, respectively. Spherical b-tensors have three identical eigenvalues and lack a well-defined direction and symmetry axis.

The effect of b-tensor shape can be made more tangible by visualizing the signal for a few commonly used shapes, using both simulated and in vivo examples. Figure 2 shows simulated signal in three diffusion tensor distributions ($P(\mathbf{D})$). In each case, the signal is calculated as the Laplace transform of the distribution of diffusion tensors

$$\mathrm{S}(\mathbf{B}) = \mathrm{S}_0 \int P(\mathbf{D}) \exp(-\mathbf{B}\!:\!\mathbf{D})\, \mathrm{d}\mathbf{D} = \langle \exp(-\mathbf{B}\!:\!\mathbf{D}) \rangle, \qquad \text{Eq. 9}$$

where the integration is over the space of symmetric positive-definite tensors (Jian et al., 2007, Westin et al., 2016). Figure 2 also shows signal measured in a healthy brain in vivo (Szczepankiewicz et al., 2019a). Even without model fitting, we can distinguish the hallmarks of microscopic anisotropy and isotropic heterogeneity as described in the figure caption. Naturally, it is the goal of signal representations and biophysical models to recover useful information about the microstructure from the observed signal, however, these efforts are not within the scope of this review but have been comprehensively described elsewhere (Jelescu and Budde, 2017, Fillard et al., 2011, Alexander, 2009, Novikov et al., 2019, Novikov et al., 2018a, Norhoj Jespersen, 2018, Nilsson et al., 2018, Assaf et al., 2019, Jelescu et al., 2020).

### *2.2. Basic gradient waveform design criteria*

A natural objective in the design of gradient waveforms for tensor-valued diffusion encoding is to minimize the encoding time that is required to achieve a given b-value and shape of the b-tensor while maintaining conditions necessary for imaging. Minimizing encoding times serves to reduce the echo time, which maximizes the signal-to-noise ratio and number of signal acquisitions per unit time. The imaging conditions generally prevent overlap between diffusion encoding gradients and imaging gradients, e.g., $\mathbf{g}(t) = 0$ during excitation, refocusing and readout. Furthermore, the residual dephasing vector must be zero at the end of the diffusion encoding to satisfy the spin-echo condition ($\mathbf{q}(\tau) = 0$, section 5.1) (Grebenkov, 2007). Finally, the selected set of b-tensors is informed by the subsequent analysis. In other words, a given methodology may require a specific distribution of b-values, rotations and shapes of b-tensors to yield estimates with optimal accuracy and precision (Chuhutin et al., 2017, Reymbaut et al., 2020, Coelho et al., 2019, Jones and Basser, 2004, Poot et al., 2009, Bates et al., 2020, Lampinen et al., 2020a).

Assuming that the imaging requirements are fulfilled, the main design goal is to maximize the encoding efficiency. Since the b-value increases with gradient amplitude and encoding time (section 2.3), we achieve the maximal b-value for any given encoding time by constantly engaging the gradients at their maximal strength during the available encoding time, using the maximal slew rate whenever gradients are switched. For linear b-tensors encoding the optimal solution is simple; gradients are constantly engaged at maximal strength along a given direction, and reversed half-way through the experiment, with adjustments made to retain the balance due to finite slew rate (Sjölund et al., 2015, Aliotta et al., 2017,

Hutter et al., 2018c). By contrast, the most efficient configuration of gradient pulses for b-tensors of rank above 1 is less straightforward. Arguably the simplest solution is to apply three orthogonal pairs of trapezoidal pulses (Mori and van Zijl, 1995) and scale the amplitude of each pair to yield arbitrary b-tensor shapes (Westin et al., 2016). Although this approach is both simple and robust, it is relatively inefficient. Instead, waveforms can be optimized in various ways to improve efficiency (Wong et al., 1995, Hargreaves et al., 2004). A numerical optimization framework for tensor-valued diffusion encoding that supports arbitrary sequence timing, hardware restrictions and b-tensor shape was presented by Sjölund et al. (2015). This framework ostensibly resolved the challenge of gradient waveform design for pure b-tensor encoding. However, as will be discussed in the coming sections, we may also want to account for additional tradeoffs imposed by physiology, effects beyond the b-tensor, and imaging artifacts.

### *2.3. Diffusion encoding efficiency*

To compare a wide range of waveform candidates, we may quantify their encoding efficiency ($\kappa$) in terms of achievable encoding strength (b-value) for a given maximal gradient amplitude ($g_{\mathrm{max}}$) and total encoding time ($t_{\mathrm{tot}}$), according to (Wong et al., 1995)

$$\kappa = \frac{4b}{\gamma^2 g_{\mathrm{max}}^2 t_{\mathrm{tot}}^3}. \qquad \text{Eq. 10}$$

For any given encoding time and maximal gradient amplitude, the waveform design with the highest $\kappa$ will yield the highest b-value and metric is scaled such that a rectangular waveform that constantly engages all axes, with zero ramp time, will give $\kappa = 100\%$ (Sjölund et al., 2015).

In Figure 3, we show the encoding efficiency of a wide range of waveform designs from literature, using the constraints defined in Figure 1 but allowing for a wide range of encoding times. Generally, the encoding efficiency is reduced as the anisotropy of the encoding is reduced, i.e., spherical encoding is less efficient than planar, and planar less than linear. Any additional constraints will necessarily reduce encoding efficiency, highlighting the need for careful consideration of the tradeoff between rapid acquisition at short echo-time versus the influence of hardware, physiology, confounders, and artifacts.

### *2.4. Scaling gradient waveforms to yield arbitrary b-tensor shapes*

Any gradient waveform that produces a b-tensor of sufficiently high rank can be rescaled to yield an arbitrary b-tensor shape with equal, or lower, rank. This fact is useful when a base-waveform is used to generate variations with different shapes (Westin et al., 2014, Westin et al., 2016) or when the timing of the sequence changes in such a way that the shape of the resulting b-tensor diverges from the desired value. In such cases, a simple adjustment of the waveform can be made to achieve any set of b-tensor eigenvalues. This is done by rotating the gradient waveform to the principal axis of the original b-tensor and scaling it by the ratio of desired and initial b-tensor eigenvalues ($\mathbf{F}_{ii} = \lambda_i'/\lambda_i$, 3×3 diagonal matrix), according to

$$\mathbf{g}'(t) = \mathbf{R}^{\mathrm{T}}\mathbf{F}\ \mathbf{R}\ \mathbf{g}(t), \quad \text{Eq. 11}$$

where $\mathbf{R}$ is the rotation matrix that brings the initial b-tensor into the principal axis system, $\mathbf{B}_{\mathrm{PAS}} = \mathbf{R}^{\mathrm{T}}\mathbf{B}\ \mathbf{R}$. For example, a subtle adjustment was made to the waveforms by Wong et al. (1995), Mori and van Zijl (1995), and Heid and Weber (1997) in Figure 1 to enforce $b_\Delta = 0$; executed at the full gradient strength on each axis, they would otherwise yield unintentionally anisotropic b-tensors.

## 3. Limitations imposed by hardware and physiology

The most critical limitations in gradient waveform design are those imposed by MRI hardware and human physiology. They are critical in the sense that they *must* be obeyed, meaning that the tradeoff between limitation and performance is a search for the maximal performance under the binary condition that the waveform is safely executable. Outlined below are limits related to the maximal gradient amplitude and slew rate, energy consumption and heating, mechanical vibrations, acoustic noise, as well as peripheral nerve stimulation.

### *3.1. Maximal gradient amplitude and slew rate*

The maximal gradient and maximal slew rate are determined by hardware components of the MRI system, related mainly to the gradient amplifier system and gradient coils. The magnetic field gradient ($g$) induced in a conductive coil is proportional to the current ($I$) in the coil, according to

$$g(t) = \eta \cdot I(t) \rightarrow g_{\max} = \eta \cdot I_{\max}, \quad \text{Eq. 12}$$

where $\eta$ is the coil efficiency, or sensitivity, defined as the field gradient strength at the origin per unit current (Hidalgo-Tobon, 2010). Here, we may assume that the coils create gradients in three orthogonal directions so that their combined effect spans three-dimensional space. From Eq. 12 we seen that the maximal gradient amplitude is proportional to the maximal current that can be delivered by the gradient amplifier. Similarly, the slew rate ($s = \mathrm{d}g/\mathrm{d}t$) is determined by the electric potential, or voltage ($U$), approximated by

$$s(t) \approx \frac{\eta}{L} \cdot U(t) \rightarrow s_{\max} \approx \frac{\eta}{L} \cdot U_{\max}, \quad \text{Eq. 13}$$

where $L$ is the inductance of the circuit. The limits on $g_{\max}$ and $s_{\max}$ are often known for a given MRI system and are straightforward design targets. Additionally, there exists two auxiliary limitations related to the 'instantaneous' and 'short' time scales. The first limit is that the instantaneous apparent power can be delivered by the gradient amplifier (Bauer et al., 2004). The apparent power, expressed as the voltage-current product, can be written in terms of the gradient and slew rate ($s(t) = \mathrm{d}g/\mathrm{d}t$) which gives an expression for the peak power

$$P(t) = U(t) \cdot I(t) \rightarrow P_{\max} \propto \max\big(g(t) \cdot s(t)\big). \quad \text{Eq. 14}$$

To capture the essence of Eq. 14, the product $g_{\max} \cdot s_{\max}$ is sometimes used as a summary metric for gradient performance (Blasche, 2017). The second requirement is that sufficient energy (or voltage) is available to the gradient amplifier system to execute each 'shot.' This short-term[5] limitation means that the gradient waveform may not draw more energy from the mains or capacitor bank than it can store and recharge between shots. The energy, calculated from the inductance and current, can be expressed as (Bauer et al., 2004)

$$E(t) = \frac{1}{2} L \cdot I^2(t) \rightarrow E_{\max} \propto |g|^2, \quad \text{Eq. 15}$$

where $|g|$ is the maximal magnitude of the gradient waveform. Although the limitations in Eqs. 14 and 15 are rarely considered, they may be the cause of failure when executing single instances of gradient waveforms that are especially taxing on the gradient system.

The sensitivity and inductance of modern gradient coils are determined by a wide range of features that are the primary concern for gradient system design, e.g., the winding density, gradient linearity, Lorentz force compensation, shielding and geometry of the coils (Turner, 1993, Blasche, 2017, Hidalgo-Tobon, 2010). Most notably, inductance is proportional to the square of the winding density ($L \propto \rho^2$), whereas the sensitivity is proportional to winding density ($\eta \propto \rho$). This means that there is always a tradeoff between maximal gradient amplitude and maximal slew rate determined by the winding density ($g_{\max} \propto \rho$ and $s_{\max} \propto \rho^{-1}$). Another major factor of the gradient performance is the size of the coil, where the overall efficiency of the system is inversely proportional to the radius of the coil to the fifth power ($\eta^2/L \propto r^{-5}$) (Turner, 1993). This relationship explains why smaller insert coils are preferable for ultra-high-performance applications whereas wide-bore systems are suited for lower gradient performance applications, although whole-body systems with ultra-high gradients exist (Setsompop et al., 2013) and are instrumental for exploring the frontier of dMRI (Jones et al., 2018).

Finally, we emphasize that maximal b-values are achieved by using all gradient axes simultaneously. This is equivalent to diffusion encoding along a vector that has a length above unity. For example, engaging gradient along x, y and z simultaneously yields a b-value that is three times higher than using a single axis, since $b \propto |\mathbf{g}|^2$ (Eq. 10) where $|\mathbf{g}|$ increases by a factor of $\sqrt{3} = |[1\ 1\ 1]|$. Note that the peak energy stored in the coils also increases by a factor of three (Eq. 15). For tensor-valued diffusion encoding, this feature can be used to promote maximal encoding efficiency by using the 'max-norm' constraint ($g_i^2 \leq g_{\mathrm{max}}^2$ for $i = \mathrm{x}, \mathrm{y}, \mathrm{z}$) or to promote arbitrary rotations by using the 'L2-norm' ($g_\mathrm{x}^2 + g_\mathrm{y}^2 + g_\mathrm{z}^2 \leq g_{\mathrm{max}}^2$), respectively (Sjölund et al., 2015). Figure 4 visualizes how rotations of waveforms inscribed within a sphere (L2-norm) and cube (max-norm) affect the amplitude requested from each gradient axis under rotations. The norm is also stated in these terms in Figure 1 in the rightmost column.

---

[5] We take the 'short term' to be on the order of a single excitation-to-readout time during which $g_{\max}$, $s_{\max}$, $P_{\max}$ and $E_{\max}$, rather than heating (3.2), are the limiting factors. Recall that the energy in Eq. 15 is stored in the gradient coil, and is largely recovered at the end of the waveform.

### *3.2. Energy consumption and heating*

Magnetic field gradients are created by driving a current in conductive coils, but electrical resistance causes this process to deposit energy in the coil material by resistive heating (Ohmic heating). Given a finite tolerance for the gradient hardware temperature, resistive heating provides another relevant limitation on gradient waveform design which is commonly referred to as the gradient systems 'duty cycle.' Although the gradient system may have a frequency dependent resistance and losses related to the gradient amplifier (Schmitt, 2013), we may approximate it by a purely resistive circuit with constant resistance in which the total electric power ($P$) is

$$P = R\, I^2, \qquad \text{Eq. 16}$$

where $R$ is the electrical resistance. Thus, executing any given gradient waveform will cause part of the power to be deposited in the gradient system and thermally coupled materials, increasing their temperature. The energy ($E$) deposited in the materials is the power integrated over time, using Eq. 12 we see that this is proportional to the integral of the gradient waveform squared

$$E \propto \int_0^\tau g(t)^2 \mathrm{d}t. \qquad \text{Eq. 17}$$

This expression shows that doubling the gradient amplitude will increase the energy consumption and heating by a factor of four and that stretching the waveform out in time scales this behavior linearly. In more practical terms for dMRI, we may relate the energy deposition for a given waveform to the b-value, such that

$$E \propto \frac{b}{t_{\mathrm{tot}}^2}, \qquad \text{Eq. 18}$$

where $t_{\mathrm{tot}}$ is the total duration of the gradient waveform. For a given b-value and waveform, the energy deposited per-shot can be effectively reduced by extending the duration, although this reduces the encoding efficiency and usually comes at a cost to the minimal echo time.

The tradeoff between encoding efficiency and energy consumption (Eq. 18) can be incorporated in the design of gradient waveforms, either by limiting the maximal gradient amplitude throughout the whole waveform, or as an additional restriction in numerical optimization (Sjölund et al., 2015). We note that a lower per-shot energy consumption yields longer echo times but may facilitate shorter repetition times due to shorter cool-down periods (Ivanov et al., 2010), possibly improving statistical precision per unit time. Furthermore, the electronical properties of gradient coils differ between the three axes (Hidalgo-Tobon, 2010). Each axis may therefore have different duty cycle limitations and react differently to rotation of the gradient waveform. The rightmost column of Figure 1 shows the energy consumed ($E_{\text{rel}}$) relative to the Stejskal-Tanner waveform. Notably, linear b-tensor encoding by SDE (Stejskal and Tanner, 1965) is the most energy efficient per $b$, and higher rank b-tensors generally require more energy.

Heating limitations in the long-term—taken to mean the time scale of the entire diffusion experiment—can be managed by modifying the order of execution. Since diffusion experiments frequently employ multiple b-values, b-tensor shapes, and rotations, the thermal load on gradient coils varies during the experiment. Since the order of the encodings can be arbitrarily rearranged without affecting the subsequent data analysis, we may optimize the sampling order to minimize peak thermal load. In demanding multi-slice experiments, gradient waveforms may even be shuffled on a per-slice level, i.e., consecutive slices can be acquired with different b-tensors (Hutter et al., 2018a). This approach also reduces the effects of temperature dependent signal bias caused by system drift (Vos et al., 2017). Moreover, total thermal load is reduced by using multiband imaging methods (Larkman et al., 2001, Breuer et al., 2005), since a single diffusion encoding is used to encode multiple slices, reducing energy-per-slice by the inverse of the acceleration factor. The acquisition of diffusion weighted data can also be interspersed by imaging sequences that are less taxing on the gradient system, such that extensions to the encoding time and cool-down periods can be avoided (Ivanov et al., 2010). Figure 5 shows the impact of sampling order on the thermal load on a hypothetical MRI system. The example shows that interleaving by volume or slice yields a markedly reduced peak thermal load as the system has time to cool between demanding waveforms.

### 3.3. *Mechanical vibration and acoustic noise*

When currents are passed through the gradient coils to produce field gradients, they interact with the magnetic field and the conductor experiences Lorentz forces. For a straight conductor in a uniform magnetic field the force per unit length is

$$\mathbf{F} = -\mathbf{B}_0 \times \mathbf{I}, \qquad \text{Eq. 19}$$

where both the magnetic field[6] and current are treated as vectors (Mansfield et al., 1994). Eq. 19 shows that the force will be proportional to the amplitude of the current and magnetic field strength. Using Eq. 12, we can loosely connect the force to a given gradient waveform as $F \propto |\mathbf{B}_0| g(t)$ (Heismann et al., 2015). Another relevant aspect of this interaction is that vibrations can be amplified at specific resonance frequencies of the structure itself (Smink et al., 2007), and the simple proportionality no longer captures the essence of the problem. Instead, the effect, or system response, at a given frequency can be estimated by investigating the overlap between the frequency response function ($R(f, \mathbf{v})$) and the

[6]Note that the main magnetic field vector, $\mathbf{B}_0$, is denoted with a subscript zero to distinguish it from the b-tensor.

Fourier transform of the gradient waveform, $g(f) = \mathcal{F}(\mathbf{v} \cdot \mathbf{g}(t))$, according to (Hedeen and Edelstein, 1997)

$$A(f) = g(f) \cdot R(f, \mathbf{v}), \quad \text{Eq. 20}$$

evaluated in a given direction **v** (unit vector). Since the gradient amplitude and current are often maximized in dMRI, we expect large forces on the MRI hardware that can potentially cause structural stress and failure. Vibrations also propagate to surrounding materials causing incoherent motion in the object (Weidlich et al., 2020), and into the air where they manifest as acoustic noise. Given the right frequency and sound pressure, acoustic noise experienced by the subject can be in excess of 110 dBA (McJury and Shellock, 2000), which can be uncomfortable or even harmful (Rosch et al., 2016).

In addition to passive hearing protection and active noise control (McJury and Shellock, 2000), the source of the mechanical stress and acoustic noise can be partially mitigated in hardware design by considering gradient coil winding and material characteristics such that the effect of Lorentz forces is minimized (Mansfield et al., 1998, Mansfield et al., 1995, Ireland et al., 2015). However, since the frequency response function can be estimated by broadband excitation of the MRI hardware (Hedeen and Edelstein, 1997), gradient waveforms can be designed to limit the power at resonance frequencies (Smink et al., 2007, Heismann et al., 2015) and thereby avoid amplification. This principle is in 'quiet MRI' (McJury and Shellock, 2000). For example, the quiet dMRI sequence proposed by Ott et al. (2015) reduces the amplitude of imaging gradients, and more importantly in this context, it reduces maximal slew rate (increased rise time) of diffusion encoding gradients and readout gradients. In a clinical setting where low acoustic noise is desired, such sequences can reduce sound pressure by 70% with negligible impact on diagnostic quality (Rosch et al., 2016). Similar modifications were also demonstrated together with multiband acceleration for more rapid acquisition (Hutter et al., 2018b). Vibrations and acoustic noise have not yet been explicitly considered in the optimization of gradient waveforms for tensor-valued diffusion encoding, however, we expect that future optimization tools will take as input the frequency response function (Eq. 20) and create waveforms that avoid significant power at resonance frequencies.

### *3.4. Peripheral nerve stimulation*

Switching of field gradients may induce currents in the subject that can cause stimulation of nerves. The stimulation can cause muscle twitching and is often uncomfortable to the subject. It can even be harmful, especially when considering stimulation of the cardiac muscle (Reilly, 1989). Therefore, it is arguably the most important aspect of gradient waveform design since it couples to patient safety and comfort. According to Faraday's law, a change in a magnetic field inside a conductive material induces an electrical current that opposes the original field change (section 5.5). For an arbitrary conductive loop in the body enclosing an area with normal vector **a**, at distance *r* from the iso-center, the induced electromotive potential ($U$) is proportional to the time derivative of the field gradient vector, and therefore related to the slew rate of the field gradient (Irnich and Schmitt, 1995, Ham et al., 1997), such that

$$U \propto r\,\mathbf{a} \cdot \frac{\mathrm{d}\mathbf{g}}{\mathrm{d}t}. \quad \text{Eq. 21}$$

The effect increases with area and distance from the iso-center and is therefore prominent in the peripheral regions of the body, hence peripheral nerve stimulation (PNS). Eq. 21 also shows that PNS increases with the size of the volume within which the gradients are linear, meaning that systems with reduced linearity (sublinear gradients) are preferable with respect to PNS which allows them to operate at higher slew-rates (Hidalgo-Tobon, 2010, Weiger et al., 2018).

The translation from electric potential to the sensation of PNS is not straightforward (Davids et al., 2019), but Eq. 21 suggests that the slew rate can be used to control the PNS response (Schulte and Noeske, 2015). It is also intuitive that the effect should become worse if the exposure to high slew rates is sustained for a longer time (Irnich and Schmitt, 1995). However, an accurate prediction may be difficult since it depends on subtle details in the waveform and subject physiology (Mansfield and Harvey, 1993, Ham et al., 1997, Tan et al., 2020).

Since tensor-valued diffusion encoding uses non-conventional gradient waveform designs, it is difficult to predict the PNS from experience or simple descriptors, i.e., a restriction on the maximal slew rate may not suffice (Reilly, 1989). Instead, peripheral nerve stimulation can be analyzed by PNS models calibrated by heuristic measurements. For example, Hebrank and Gebhardt (2000) modeled PNS in terms of a resistive-capacitive circuit to predict PNS for arbitrary gradient waveforms, and more recent models have been proposed to take into account the intricacies of human physiology (Davids et al., 2019).

In Figure 6, we use the SAFE model by Hebrank and Gebhardt (2000) to estimate PNS levels caused by several diffusion encoding waveforms. We compare them with respect to peak and cumulative PNS assuming a hypothetical MRI system. The figure highlights that even if all waveform designs are constrained to the same maximal slew rate, they cause variable levels of PNS because they have different gradient and slewing trajectories.

## 4. Sensitivity to features beyond Gaussian diffusion

By describing the experiment with the b-tensor alone, we implicitly state that our interest lies in capturing the behavior of multi-Gaussian diffusion. However, depending on the microstructure and physiology of the investigated tissue, diffusion encoding may sensitize the signal to features beyond Gaussian diffusion which renders the b-tensor description incomplete. If the influence of such features on signal is not negligible, they may either be harnessed to probe more intricate tissue details, or act as confounders. This distinction motivates the current naming convention, where ‘spherical b-tensor encoding’ is preferred over 'isotropic diffusion encoding' precisely because the *b-tensor* is designed to be spherical whereas other effects related to diffusion may be rotation variant[7]. In this section, we highlight three features that are not captured by the b-tensor formalism, but may still inform the gradient waveform design, namely effects of timing in the context of restricted diffusion, exchange, and incoherent motion.

### *4.1. Time-dependent diffusion due to restriction*

Diffusion MRI in biological tissue relies on microscopic hindrances and restrictions to produce interesting contrast. In the presence of such structures, the assumption that the diffusion process is approximately Gaussian may be inaccurate (Stepisnik, 1999). Consider, for example, diffusion in a closed compartment. If diffusion is observed during a short time, the water does not have time to probe its confinement and the apparent diffusivity approaches that of free diffusion (Mitra et al., 1993, Stepisnik, 1993). By contrast, a long time allows the water to sense the entire confinement, particles are trapped near their origin, and the apparent diffusivity approaches zero. If they are merely hindered, the apparent diffusivity will be some finite value related to the geometric configuration of obstacles (Novikov et al., 2014). In the extreme cases of very short or very long diffusion times, the diffusion process can be well described by Gaussian diffusion, however, in the intermediate regime, there is an interplay between the geometry of the restriction and the diffusion time (Stepisnik, 1993). Since the diffusion time is related to the diffusion encoding waveform, we may design it to be more or less sensitive to diffusion time effects. This relationship provides a rich contrast that has been studied with specialized waveform designs to probe size and shape of restrictions (Callaghan and Komlosh, 2002, Portnoy et al., 2013, Aggarwal et al., 2012, Colvin et al., 2008, Does et al., 2003, Reynaud et al., 2016, Gore et al., 2010, Lemberskiy et al., 2017, Özarslan and Basser, 2008, Laun and Kuder, 2013, Reynaud, 2017, Sen, 2004, Mitra et al., 1993, Stepisnik, 1993, Henriques et al., 2020, Ianus et al., 2018, Lundell et al., 2019).

It is important to realize that all gradient waveforms can be sensitive to diffusion time effects (Stepisnik, 1993). However, if diffusion time effects are not explicitly sought after, the tendency is for waveforms to be designed for maximal encoding efficiency which indirectly promotes long diffusion times. Additionally, encoding along a single direction per acquisition generally uses the same waveform for all measurement. This means that even if diffusion time effects are unknown, they are consistent across all measurements. This provides a rotation invariant 'snapshot' of the diffusion process; a condition that is widely accepted as a necessary compromise to yield simple dMRI experiments. Since this may not be true for intricate waveforms for tensor-valued diffusion encoding, a waveform and rotation dependent confounder may be introduced.

---

[7] This is congruent with 'weighting by the trace of the diffusion tensor' as used by Mori and van Zijl (1995) and Heid and Weber (1997). Naturally, this naming convention extends to other shapes of the b-tensor.

This problem was pointed out by de Swiet and Mitra (1996) soon after the introduction of intricate waveforms for isotropic diffusion encoding. Here, we distinguish two aspects of this problem that may inform gradient waveform design. First, gradient waveforms designed to produce b-tensors of different shapes may have different overall diffusion time characteristics (between waveform differences). Second, waveforms that span multiple dimensions ($\text{rank}(\mathbf{B}) > 1$) may have diffusion time characteristics that differ along different directions (rotation variance or spectral anisotropy). For example, linear encoding by SDE generally exhibits a longer diffusion time compared to waveforms that yield spherical b-tensor encoding which tend to comprise oscillating patterns. The two waveforms may therefore yield different estimates of diffusion parameters where the Gaussian assumption would predict them to be equal (Lundell et al., 2019).

Although the impact of diffusion time on tensor-valued diffusion encoding is still being explored across waveform designs and tissues (Jespersen et al., 2019, Szczepankiewicz et al., 2019b, Clark et al., 2001, Lundell et al., 2019, Nilsson et al., 2017), a potential solution to avoid the confounder is to design waveforms that are matched with respect to diffusion times. Doing so would restore the 'snapshot' condition for all variants of the b-tensor (Lundell et al., 2020). Alternatively, if the effects of diffusion time are non-negligible, the diffusion time can be purposefully varied and included in a more comprehensive quantification (Drobnjak and Alexander, 2011, Nilsson et al., 2020b, Henriques et al., 2020).

Diffusion time characteristics of arbitrary gradient waveforms can be analyzed in terms of the diffusion encoding power spectrum. For simplicity, we use a one-dimensional representation by describing the effect along a single direction $\mathbf{v}$ (unit vector) at a time, such that $q(f) = \mathcal{F}\left(\gamma \int_0^t g(t)\mathrm{d}t'\right)$ is the Fourier transform of the dephasing vector (Eq. 6) where $g(t) = \mathbf{v} \cdot \mathbf{g}(t)$. The diffusion weighted signal can be approximated by the product between the encoding power spectrum ($|q(f)|^2$) and the diffusion spectrum ($D(f)$) which is the Fourier transform of the velocity autocorrelation of diffusion particles (Stepisnik, 1993), according to

$$S \approx \exp\left(-\int_{-\infty}^{\infty} D(f) \cdot |q(f)|^2 \, \mathrm{d}f\right). \qquad \text{Eq. 22}$$

Two waveforms are matched with respect to diffusion time characteristics if their encoding spectra are similar. In sufficiently small structures, we may simplify this condition and capture the relevant aspects of the diffusion time characteristics with the variance of the encoding power spectrum ($V_\omega$), defined as (Nilsson et al., 2017)

$$bV_\omega = \gamma^2 \int_0^\tau g^2(t) \, \mathrm{d}t, \qquad \text{Eq. 23}$$

provided that their size is below a threshold, loosely defined as $\sqrt{D_0/f_0}$, where $D_0$ is the bulk diffusivity, and $f_0$ is the highest frequency at which the encoding spectrum has relevant power. Preliminary efforts have also extend this simplified analysis to three dimensions (Lundell et al., 2018) and an in depth treatment of spectral anisotropy was presented by Lundell and Lasič (2020). To the best of our knowledge, there are currently no tools for optimizing waveforms for tensor-valued diffusion encoding that also consider diffusion time matching, however, we expect encoding efficiency for such waveforms to be limited due to the unfavorable relationship between b-value and encoding time.

To make the effects of diffusion time more tangible, we visualize the impact of diffusion time on the apparent diffusivity for multiple waveform designs in Figure 7. The apparent diffusion coefficient is calculated for a one-dimensional stick of finite length (Stepisnik, 1993) for a comprehensive set of stick directions, such that the stick experiences all projections (or rotations) of the waveform. This geometry was chosen to maximize the effect of diffusion time effects to depict a 'worst case' scenario. The figure highlights that different waveforms may yield different average apparent diffusivity (between waveform differences), different levels of rotation variance (within waveform spectral anisotropy), but that there also exist length scales at which diffusion time effects are negligible (Nilsson et al., 2017, Grebenkov, 2007).

### *4.2. Time-dependent diffusion due to exchange*

During a single preparation of the signal, water molecules may traverse across permeable boundaries and thereby undergo compartment exchange. By this process, water may visit environments associated with different apparent diffusivities (Kärger, 1985). For example, consider exchange over the cell membranes that separate the intra and extra-cellular space. The sensitivity of the diffusion weighting to this process depends on the shape of the gradient waveform. Methods such as filter exchange imaging employ pairs of bipolar pulses separated by a waiting time to encode the exchange rate in order to probe the permeability of compartment boundaries in biological tissue (Eriksson et al., 2017, Benjamini et al., 2017, Lampinen et al., 2016, Lasič et al., 2011).

Analogous to diffusion time effects caused by restriction, the exchange weighting associated with waveforms used for tensor-valued diffusion encoding is rotation variant and differs across waveform designs. Ning et al. (2018) proposed a definition for the exchange weighting time ($\Gamma$) for arbitrary gradient waveforms, according to

$$\Gamma = \frac{2}{b^2} \int_0^\tau q_4(t)\, t \, \mathrm{d}t, \qquad \text{Eq. 24}$$

where $q_4(t) = \int_0^\tau q^2(t')q^2(t'+t)\,\mathrm{d}t'$ and $q(t)$ is the dephasing along a direction vector $\mathbf{v}$ (unit vector), such that $q(t) = \gamma \int_0^t \mathbf{v} \cdot \mathbf{g}(t')\mathrm{d}t'$. We note that large values of $\Gamma$ indicate that the exchange process is given a longer time to act, making the experiment more sensitive to exchange and that waveforms with similar values of $\Gamma$ will produce similar effects of exchange.

We visualize the rotation variance of exchange weighting time in Figure 7 by calculating $\Gamma$ along multiple directions $\mathbf{v}$. Although the figure does not show how exchange weighting is translated into observed signal, it illustrates that exchange weighting times vary across waveform designs and rotations.

### *4.3. Incoherent motion*

In addition to the diffusion process driven by thermal motion, spin dephasing is caused by other kinds of incoherent motion. Most notably, the relatively slow and incoherent flow of blood in capillaries has a measurable impact on the diffusion weighted signal at low b-values and carries information about the vasculature which can be mistaken for fast diffusion, also called 'pseudo diffusion' (Le Bihan et al., 1986, Ahn et al., 1987). Other sources of motion include cardiac and pulmonary motion, which may influence diffusion measurements in brain by arterial pulsation (Skare and Andersson, 2001, Habib et al., 2010) and by gross movement of tissue, for example in chest, cardiac and kidney imaging (Haacke and Lenz, 1987, Lasic et al., 2020, Nery et al., 2019), or from vibrations induced by the diffusion encoding itself (Weidlich et al., 2020, Gallichan et al., 2010, Hiltunen et al., 2006) as described in section 3.3.

Incoherent motion within a voxel causes dephasing related to the motion encoding strength of the gradient waveform (Stejskal, 1965, Callaghan and Stepišnik, 1996). The motion encoding vector of arbitrary order ($n$) can be computed as

$$\mathbf{m}_n = \gamma \int_0^\tau \mathbf{g}(t)\, t^n \,\mathrm{d}t, \qquad \text{Eq. 25}$$

where $\mathbf{m}_0$ encodes position and is always zero to satisfy the spin-echo condition (section 5.1), $\mathbf{m}_1$ encodes velocity, $\mathbf{m}_2$ encodes acceleration and so on. Curiously, velocity encoding can also be gleaned from the encoding power spectrum (section 4.1), where the power at zero frequency corresponds to velocity encoding[8]. Unlike diffusion time and exchange, motion encoding is always vector-valued, and therefore exhibits a single direction along which it is highest and is zero in orthogonal directions. Thus, the only way to make a waveform with isotropic motion encoding is to make it motion compensated. For symmetric and self-balanced gradient waveforms ($\mathbf{q}(t) = 0$ during the refocusing pulse), velocity compensation can be simply turned on/off by reversing the direction of the encoding waveforms on either side of the refocusing. For example, two pairs of bipolar pulses can be configured in a parallel or anti-parallel fashion to yield encoding that is compensated or non-compensated for velocity (Ozaki et al., 2013, Ahlgren et al., 2016). More generally, we may exploit special symmetries to gain compensation of arbitrary moments (Pipe and Chenevert, 1991), which is applicable to tensor-valued encoding

[8] Using integration by parts, it can be shown that $\mathbf{m}_1 = \gamma \int_0^\tau \mathbf{g}(t)\, t \,\mathrm{d}t = -\int_0^\tau \mathbf{q}(t)\, \mathrm{d}t$ for balanced gradient waveforms ($\mathbf{q}(\tau) = 0$). Thus, the power spectrum of $\mathbf{q}$ at zero frequency is related to velocity encoding because $\mathbf{q}(f = 0) = \int_0^\tau \mathbf{q}(t)\, \mathrm{d}t = -\mathbf{m}_1$.

(Lasic et al., 2020). Motion compensation can also be achieved by optimizing gradient waveforms while constraining the magnitude of the $n^{\text{th}}$ moment vector to a threshold value ($L$) that ensures negligible motion encoding

$$|\mathbf{m}_n| \leq L_n. \quad \text{Eq. 26}$$

Based on this principle, an optimization framework for motion compensated waveforms was proposed for linear encoding (Aliotta et al., 2017) and a similar approach was recently developed for tensor-valued diffusion encoding (Szczepankiewicz et al., 2020b). Naturally, there exists a tradeoff between constraining higher order moments and encoding efficiency, where constraints on motion encoding for order 0 to $n$ generally reduces the encoding efficiency as $n$ increases (compare motion compensated waveforms in Figure 3). Therefore, alternative methods to alleviate effects of incoherent motion may be preferred. For example, the effect of capillary blood flow can be suppressed by avoiding sampling at low b-values (Le Bihan, 2013), systematic motion can be avoided by using gated imaging (Nunes et al., 2005), and post-processing can alleviate pulsation artefacts (Skare and Andersson, 2001, Gallichan et al., 2010).

In the rightmost column of Figure 7, velocity encoding is shown for several gradient waveforms. As with diffusion time and exchange, the strength of velocity encoding can vary between waveform designs and across rotations of any given waveform. However, designs for motion compensated tensor-valued diffusion encoding are readily available to ensure control of this contrast or potential confounder (Lasic et al., 2020, Szczepankiewicz et al., 2020b).

## 5. Artifacts and errors related to diffusion encoding gradients

Several imaging artifacts and sources of signal error are related directly to the diffusion encoding gradient waveform. Unlike the potential confounders in the previous section, these features are unlikely candidates for biomarkers, and their removal will be considered a categorical improvement of the experiment. Here, we survey the effects of gradient balance, concomitant gradients, non-linear gradients, background gradients, and eddy-currents.

### *5.1. Gradient balance and residual dephasing*

Diffusion encoding gradient waveforms are designed to be 'balanced,' i.e., they are intended to have a negligible $0^{\text{th}}$ moment such that no phase change is induced to any stationary spin (Grebenkov, 2007). In other words, the dephasing vector, $\mathbf{q}(t)$, should not contribute any residual dephasing to the readout trajectory, $\mathbf{k}(t)$. Note that these share the same definition, but are denoted differently for historical reasons, and to emphasize their difference in function (Blümich, 2016). We may capture the imbalance in terms of a residual moment at the end of diffusion encoding, according to

$$\tilde{\mathbf{k}} = \gamma \int_0^{t_{\text{tot}}} \mathbf{g}(t)\mathrm{d}t = \mathbf{q}(t_{\text{tot}}), \qquad \text{Eq. 27}$$

where the gradient waveform, $\mathbf{g}(t)$, is said to be balanced only if $\tilde{\mathbf{k}} \approx 0$. This condition is normally straightforward to satisfy, e.g., by using a 'symmetric' gradient waveform that is identical before and after the refocusing pulse such that the sign reversal of the refocusing pulse ensures that the moment before and after are equal and opposite. Conveniently, any errors that depend only on the gradient waveform will also be symmetric, and therefore preserve the spin-echo condition. By contrast, asymmetric gradient waveforms forego the robustness from symmetry in favor of efficiency (section 2.3), and additional care is therefore required to ensure balance. For example, a residual moment ($\tilde{\mathbf{k}} \neq 0$) can appear due to effects of interpolation and re-sampling of waveforms, inaccurate duration and/or concomitant gradients (section 5.2).

The effects of residual moments can be challenging to detect visually because they depend not only on the magnitude of $\tilde{\mathbf{k}}$, but also its orientation with respect to the imaging and readout gradients. For example, in an ideal spin-echo sequence with rectangular slice profiles and two-dimensional echo-planar imaging readout (Mansfield, 1977), the signal bias factor (*BF*) (Menditton et al., 2006) caused by a residual dephasing can be approximated by (Baron et al., 2012, Norris and Hutchison, 1990, Du et al., 2002, Szczepankiewicz et al., 2019d)

$$BF \approx \left|\text{sinc}\left(\mathbf{n}_\text{s} \cdot \tilde{\mathbf{k}} \cdot ST\right)\right| \cdot \exp\left(-\left|\mathbf{n}_\text{p} \cdot \tilde{\mathbf{k}}\right| \frac{\Delta t}{\Delta k \cdot T_2^*}\right), \qquad \text{Eq. 28}$$

where $\mathbf{n}_\text{s}$ and $\mathbf{n}_\text{p}$ are unit vectors pointing along the slice normal and phase-encoding directions, $ST$ is the slice thickness, $\Delta t$ is the echo-spacing time, $\Delta k$ is the step size in k-space, and $T_2^*$ is the observed transversal relaxation rate. Importantly, $\tilde{\mathbf{k}} \propto \sqrt{b}$, and the loss of signal can therefore be mistaken for diffusion effects, leading to gross misestimation of microstructure parameters and fiber orientations (Baron et al., 2012, Szczepankiewicz et al., 2019d).

In cases where $\tilde{\mathbf{k}}$ is relatively small and uniform, it can be nulled by adding a 'balancing gradient' in the pulse sequence that has a $0^{\text{th}}$ moment with equal magnitude but opposite direction ($\mathbf{k}_{\text{bal}} = -\tilde{\mathbf{k}}$) (Szczepankiewicz, 2016).

### *5.2. Concomitant gradients*

Residual gradient moments, as described in section 5.1, may also be caused by so-called 'concomitant gradients' or 'Maxwell terms.' Concomitant gradients are always created alongside the desired gradient waveform and can cause severe image artifacts and signal dropout (Eq. 27). The concomitant gradient waveform can be approximated by (Szczepankiewicz et al., 2019d, Baron et al., 2012, Meier et al., 2008, Bernstein et al., 1998)

$$\mathbf{g}_c(t,\mathbf{r}) \approx \frac{1}{4|\mathbf{B}_0|}\begin{bmatrix} g_z^2(t) & 0 & -2g_x(t)g_z(t) \\ 0 & g_z^2(t) & -2g_y(t)g_z(t) \\ -2g_x(t)g_z(t) & -2g_y(t)g_z(t) & 4g_x^2(t)+4g_y^2(t) \end{bmatrix} \cdot \mathbf{r}\, h(t), \qquad \text{Eq. 29}$$

where $|\mathbf{B}_0|$ is the main magnetic field strength and $\mathbf{r}$ is the position vector relative to the isocenter[9]. Concomitant gradients are generally too weak to have a relevant effect on the actual diffusion encoding b-tensor (Baron et al., 2012), but they may contribute to a position-dependent dephasing vector at the end of the encoding ($\tilde{\mathbf{k}}(\mathbf{r}) \neq 0$, Eq. 27), which is no longer straightforwardly removable by a linear balancing gradient. This effect is especially prominent in dMRI due to the high gradient magnitudes used for diffusion encoding, but can also be caused by imaging gradients (Irfanoglu et al., 2012). The dephasing vector depends on the amplitude and rotation of the gradient waveform, and its impact on the signal depends on multiple imaging parameters (voxel size, bandwidth, etc.) as described in Eq. 28. It is therefore difficult to detect and recognize in raw data (Zhou et al., 1998, Du et al., 2002) and can easily be mistaken for diffusion effects (Baron et al., 2012), in turn causing gross errors in subsequent quantification (Szczepankiewicz et al., 2019d).

For linear diffusion encoding the effects of concomitant gradients can be compensated by ensuring that the integral of the square of the gradient waveform in a spin-echo sequence is equal before and after the refocusing pulse (Zhou et al., 1998). Errors can also be suppressed around a single point in space by an online subtraction of the predicted concomitant gradients (Meier et al., 2008). For tensor-valued encoding, compensation can be achieved by optimizing waveforms where the so called 'Maxwell index' ($m$) is constrained to a sufficiently small value, also known as 'M-nulling.' The Maxwell index is defined as (Szczepankiewicz et al., 2019d)

$$m = \sqrt{\mathrm{Tr}(\mathbf{M}\mathbf{M})}, \qquad \text{Eq. 30}$$

where $\mathbf{M} = \int_0^\tau h(t)\mathbf{g}(t)\otimes\mathbf{g}(t)\mathrm{d}t$ and $h(t)$ is a sign function that assumes values of ±1 to indicate the direction of spin dephasing. Waveforms with sufficiently low values of $m$ will have concomitant gradients with negligible residual 0$^{\text{th}}$ moment vectors ($\tilde{\mathbf{k}} \approx 0$, Eq. 27), independent of scaling and rotation of the waveform, and position within the FOV.

An alternative strategy is to ensure that the matrix in Eq. 29 integrates to zero, or so called 'K-nulling' (Szczepankiewicz et al., 2019d, Lasič et al., 2020). Although this approach yields gradient waveforms with slightly higher efficiency, they cannot be arbitrarily rotated without compromising the concomitant gradient compensation, nor is the compensation robust to gradient non-linearity (Szczepankiewicz et al., 2020a). Nevertheless, K-nulling may be preferred in applications where rotation of the waveform is not necessary, such as spherical b-tensor encoding, preferably in conjunction with max-norm optimization.

[9] Unlike the formalism in Szczepankiewicz et al. (2019d), here $\mathbf{g}_c(t,\mathbf{r})$ is defined in terms of the 'effective gradient,' such that $\gamma \int_0^\tau \mathbf{g}_c(t,\mathbf{r})dt \approx 0$ indicates a balanced and 'Maxwell compensated' waveform. See Eq. 30 for description of $h(t)$.

Figure 8 shows examples of several gradient waveforms, their concomitant gradients, and their impact on the final dephasing vector. It showcases that concomitant gradients can cause a complete loss of signal, even at moderate b-values and clinical MRI hardware specifications (Szczepankiewicz et al., 2019d). Although many waveform designs throughout literature exhibit this issue, a particularly common case is DDE in a spin-echo where the bipolar pairs are not co-linear and placed on either side of the refocusing pulse, as shown in Figure 8. A solution to this problem was presented by Wedeen et al. (2006) who modified a DDE waveform to be symmetric in a spin echo—essentially using negative mixing times (Shemesh et al., 2016)—thereby gaining more efficient planar diffusion encoding for improved tractography Wedeen et al. (2012) without concomitant gradient effects. Remarkably, even with the additional constraint on the Maxwell index, numerical optimization produces yet more efficient waveforms that are free from concomitant gradient effects (Szczepankiewicz et al., 2019d).

### *5.3. Non-linear gradients*

The magnetic field gradient is generally assumed to be linear within the maximal field of view of any given MRI system. However, due to practical considerations in the design of the gradient coils (Hidalgo-Tobon, 2010), the field gradient may diverge from linearity; referred to as 'gradient non-linearity' or 'non-uniformity.' Non-linearity can be measured for each MRI system and subsequently corrected for during image reconstruction (Doran et al., 2005) and analysis of diffusion-weighted data (Bammer et al., 2003). The non-linearity can be modeled by a linear transform ($\mathbf{L}$, 3×3 matrix) at positions ($\mathbf{r}$) in space such that the actual gradient waveform—perturbed by gradient non-linearity—is approximated by (Bammer et al., 2003)

$$\mathbf{g}_{\mathrm{gnl}}(t, \mathbf{r}) = \mathbf{L}(\mathbf{r}) \cdot \mathbf{g}(t). \qquad \text{Eq. 31}$$

If $\mathbf{L}$ is known throughout the imaging volume, an accurate position dependent b-tensor can be calculated by inserting $\mathbf{g}_{\mathrm{gnl}}$ in Eq. 5 and Eq. 6, such that $\mathbf{B}_{\mathrm{gnl}}(\mathbf{r}) = \mathbf{L}(\mathbf{r})\, \mathbf{B}\, \mathbf{L}^{\mathrm{T}}(\mathbf{r})$. These spatially varying b-tensors can be used in the subsequent analysis to recover accuracy (Tan et al., 2013, Jovicich et al., 2006, Rudrapatna et al., 2020).

Due to the tradeoff between gradient performance and gradient linearity (Hidalgo-Tobon, 2010), effects of non-linearity are likely most pronounced in systems designed to deliver ultra-high field strength (Mesri et al., 2019, Bammer et al., 2003) and for insert coils with compact designs. Gradient non-linearity also affects the concomitant field gradients. For example, Maxwell compensation by M-nulling (section 5.2) is robust to non-linearity, whereas K-nulling is not (Szczepankiewicz et al., 2020a).

### 5.4. Background gradients

In any inhomogeneous magnetic field, there exist so-called 'intrinsic' or 'background' field gradients. Background gradients are caused by susceptibility differences between tissues or poor shimming, and can extend over macroscopic and microscopic scales, i.e., on the scale of voxels or on the sub-voxel level (Zheng and Price, 2007, Novikov et al., 2018b). For readout methods that have a limited per-voxel bandwidth, such as EPI, regions with pronounced field inhomogeneity exhibit geometric distortion, also referred to as susceptibility artifacts (Jezzard and Balaban, 1995). In addition to imaging artifacts, background gradients create their own diffusion encoding, as well as 'cross-terms' with the desired gradient waveform, both of which introduce a bias to the desired diffusion weighting (Zheng and Price, 2007, Lawrenz and Finsterbusch, 2019, Pampel et al., 2010).

We may treat the background gradient as an unknown gradient waveform ($\mathbf{g}_\mathrm{b}(t)$) which will be added to the desired gradient waveform ($\mathbf{g}_\mathrm{d}(t)$). When we replace the gradient waveform in equation Eq. 6 with the sum of the desired and background gradient waveform (Stejskal and Tanner, 1965), the actual b-tensor from Eq. 5 can be written as

$$\begin{aligned}\mathbf{B} = \int_0^\tau \mathbf{q}_\mathrm{d}(t)\otimes\mathbf{q}_\mathrm{d}(t) + \underbrace{\mathbf{q}_\mathrm{d}(t)\otimes\mathbf{q}_\mathrm{b}(t)+\mathbf{q}_\mathrm{b}(t)\otimes\mathbf{q}_\mathrm{d}(t)}_{\text{cross-terms}} + \mathbf{q}_\mathrm{b}(t)\otimes\mathbf{q}_\mathrm{b}(t)\,\mathrm{d}t \\ = \mathbf{B}_\mathrm{d} + \mathbf{B}_\mathrm{c} + \mathbf{B}_\mathrm{b},\end{aligned} \qquad \text{Eq. 32}$$

where $\mathbf{B}_\mathrm{c}$ and $\mathbf{B}_\mathrm{b}$ are contributions to the b-tensor from cross-terms and pure background gradients, respectively. Assuming that the background gradient is approximately stationary in time[10], with an amplitude ($|\mathbf{g}_\mathrm{b}|$) and direction ($\mathbf{n}_\mathrm{b}$), we can write it as

$$\mathbf{g}_\mathrm{b}(t) = |\mathbf{g}_\mathrm{b}|\mathbf{n}_\mathrm{b}h(t), \qquad \text{Eq. 33}$$

where $h(t)$ is the spin dephasing sign function (see description of Eq. 30). Since a refocusing pulse changes the effective sign of the background gradient by switching the value of $h(t)$, the diffusion encoding effect of $\mathbf{g}_\mathrm{b}(t)$ can be modulated or nulled by the RF design (Karlicek and Lowe, 1980, Lian et al., 1994). Alternatively, $\mathbf{B}_\mathrm{b}$ can be estimated explicitly for any given imaging setup, and incorporated in the analysis to remove its confounding effect (Jara and Wehrli, 1994, Lian et al., 1994). In human dMRI, the challenge of background gradients can be traced mainly to the cross-term, i.e., $\mathbf{B}_\mathrm{c}$ is relatively large compared to $\mathbf{B}_\mathrm{b}$. Since the cross-term depends on the desired gradient waveform, there is an opportunity to designs waveforms that minimize the impact of cross-terms on measurement accuracy. For example, Hong and Dixon (1992) configured a DDE waveform to be symmetric about the refocusing pulse in a spin-echo to yield linear b-tensor encoding with negligible cross-terms. A similar symmetry was used by de Graaf et al. (2001) where bipolar gradient pairs were configured around two refocusing pulses to yield spherical b-tensor encoding while cross-terms were eliminated, later extended to also account for cross-terms with imaging gradients (Valette et al., 2012). An efficient and flexible waveform design for linear b-tensor encoding was proposed by Finsterbusch (2008) where cross-terms could be suppressed for arbitrary sequence timing, thereby making it feasible for in vivo imaging.

[10] The relevant time scale is defined by the duration of the diffusion encoding in a single experiment. This means that spin must experience an approximately stationary background gradient on time scales on the order of 10-100 ms.

To gauge the 'cross-term sensitivity' of an arbitrary desired gradient waveform, we propose the metric

$$\mathbf{c} = \int_0^\tau \mathbf{q}_\mathrm{d}(t) H(t)\, \mathrm{d}t, \qquad \text{Eq. 34}$$

where $H(t) = \int_0^t h(t')\, \mathrm{d}t'$. For background gradients that are approximately stationary with arbitrary amplitude and direction, any waveform with $\mathbf{c} = 0$—meaning that $\mathbf{q}_\mathrm{d}(t)$ and $\mathbf{q}_\mathrm{b}(t)$ are orthogonal—is unaffected by cross-terms ($\mathbf{B}_\mathrm{c} = 0$) and can be said to be 'cross-term-nulled.' Indeed, tensor-valued diffusion encoding with such cross-term-nulling can potentially be achieved by constraining $|\mathbf{c}|$ during waveform optimization such that $\mathbf{B}_\mathrm{c}$ is suppressed to negligible values. However, in the presence of severe enough background gradients, $\mathbf{B}_\mathrm{b}$ will not be negligible and may cause a bias in the actual b-tensor and subsequent errors in quantification.

In lieu of optimized waveforms with cross-term-nulling, two simple approaches exist for cross-term suppression. For a spin-echo sequence, where $H(t)$ is a triangular function centered on $\tau/2$, Eq. 34 states that **c** is zero for any effective gradient waveform that is mirrored in the refocusing pulse, i.e., the second part is the time-reverse of the first part ($\mathbf{g}(t) = \mathbf{g}(\tau - t)$). For example, we can achieve $\mathbf{c} = 0$ and $\tilde{\mathbf{k}} = 0$ (Eq. 27) by executing any self-balanced gradient waveform (section 4.3) before and after the refocusing pulse, where the second execution reverses both the time and polarity compared to the first (Hong and Dixon, 1992). Although this approach is simple, it relies on self-balanced waveforms which have poor encoding efficiency (Figure 3). Alternatively, any waveform can be used to perform the measurement twice, each time with opposite gradient polarity ($\mathbf{g}(t)$ and $-\mathbf{g}(t)$). Thereafter, the geometric average of the two signals can be computed, and will be independent of cross-terms (Neeman et al., 1991) and therefore behave like measurements performed with a waveform with negligible cross-term sensitivity.

In Figure 9, we visualize the manifestation of background gradients for a set of waveform variants based on double diffusion encoding in a spin-echo, including a cross-term-nulled design and repeated acquisition with reversed gradient polarity.

### 5.5. *Eddy currents*

As for the conductive materials of the body, rapid changes in the gradient waveform also induce currents in the conductive materials of the MRI scanner. Such currents are referred to as eddy currents. In addition to heating the conductive material, eddy currents generate magnetic fields which counteract the execution of the gradient waveform that caused them. Since eddy current have finite decay times, they may linger and perturb the imaging gradients of the readout, causing image artifacts (Jezzard et al., 1998).

The temporal behavior of eddy current field gradients ($\mathbf{g}_{\mathrm{ec}}$) can be approximated by a convolution of the time-derivative of the desired gradient waveform ($\mathrm{d}\mathbf{g}/\mathrm{d}t$) with a set of truncated exponential impulse functions ($F(t)$ is the Heaviside function) with decay constants ($\lambda_i$) specific to the system, such that (van Vaals and Bergman, 1990, Bernstein et al., 2004)

$$\mathbf{g}_{\mathrm{ec}}(t) = -\frac{\mathrm{d}\mathbf{g}(t)}{\mathrm{d}t} \circledast \sum_i w(\lambda_i) F(t) \exp\left(-\frac{t}{\lambda_i}\right), \qquad \text{Eq. 35}$$

where $w_i$ are system specific amplitudes and the operator $\circledast$ denotes convolution. In principle, the amplitudes and time constants can be calculated from the characteristics of the MRI system, but are more frequently measured directly (Bernstein et al., 2004). Notably, eddy current artifacts will depend on the direction and strength of the diffusion encoding (Jezzard et al., 1998), making them heterogeneous across the dMRI experiment.

Eddy currents are especially prominent for the gradient waveforms used for diffusion encoding since these are generally much stronger than those used for imaging. Since eddy currents oppose the desired application of gradients, the waveform shape is itself distorted. However, the desired shape can be retained by so called gradient ‘pre-emphasis correction,’ where the effect of eddy currents is predicted and accounted for during the execution of a waveform (Bernstein et al., 2004). Eddy current effects can also be considered explicitly in the design of the encoding waveform. Approaches include eddy current compensation by using bi-polar pulses (Alexander et al., 1997), adjusting the timing of asymmetric pulses in a double spin-echo sequence (Reese et al., 2003, Finsterbusch, 2010), and a general optimization framework for eddy current minimization for arbitrary waveform timing and RF-pulse setup (Aliotta et al., 2018). Eddy currents have also been accounted for in numerical optimization of tensor-valued diffusion encoding by Yang and McNab (2018), based on the framework of Sjölund et al. (2015). We note that waveforms designed to minimize eddy currents will likely reduce the efficiency of the diffusion encoding, whereas a post-processing approach will not (Irfanoglu et al., 2019, Nilsson et al., 2015), calling for a careful consideration of how such effects should be handled.

## 6. Concluding remarks

Tensor-valued diffusion encoding has proven its value as part of the dMRI toolbox, however, the design of the gradient waveforms used for encoding still presents several challenges. Although many such challenges have straightforward solutions, the designer of the experiment must carefully consider the tradeoff between speed of acquisition and data quality versus confounding effects and artifacts which may compromise the accuracy of both quantification and interpretation.

Given the direction of ongoing activities, we foresee that dMRI experiments will come to consider all relevant features in the design of waveforms, pulse sequences and sampling schemes. Doing so will allow us to harness the benefits of multidimensional correlation-experiments to facilitate joint estimation of parameters related to diffusion, restriction, exchange, motion, relaxation and beyond.

### CRediT authorship contribution statement

**Filip Szczepankiewicz:** Conceptualization, Investigation, Analysis, Software, Visualization, Writing - Original Draft, Writing - Review & Editing, Supervision. **Carl-Fredrik Westin:** Conceptualization, Writing - Review & Editing, Funding Acquisition. **Markus Nilsson:** Conceptualization, Writing - Review & Editing, Software, Funding Acquisition.

### Acknowledgements

This study was supported by the Swedish Research Council, Sweden, grant no. 2016-03443, and the NIH, United States, grant no. R01MH074794 and P41EB015902.

### Declaration of competing interest

FS and MN are inventors on patents related to gradient waveform design. MN holds shares in a company that holds intellectual property in methods related to dMRI. CFW has no competing interests to declare.

## Figures

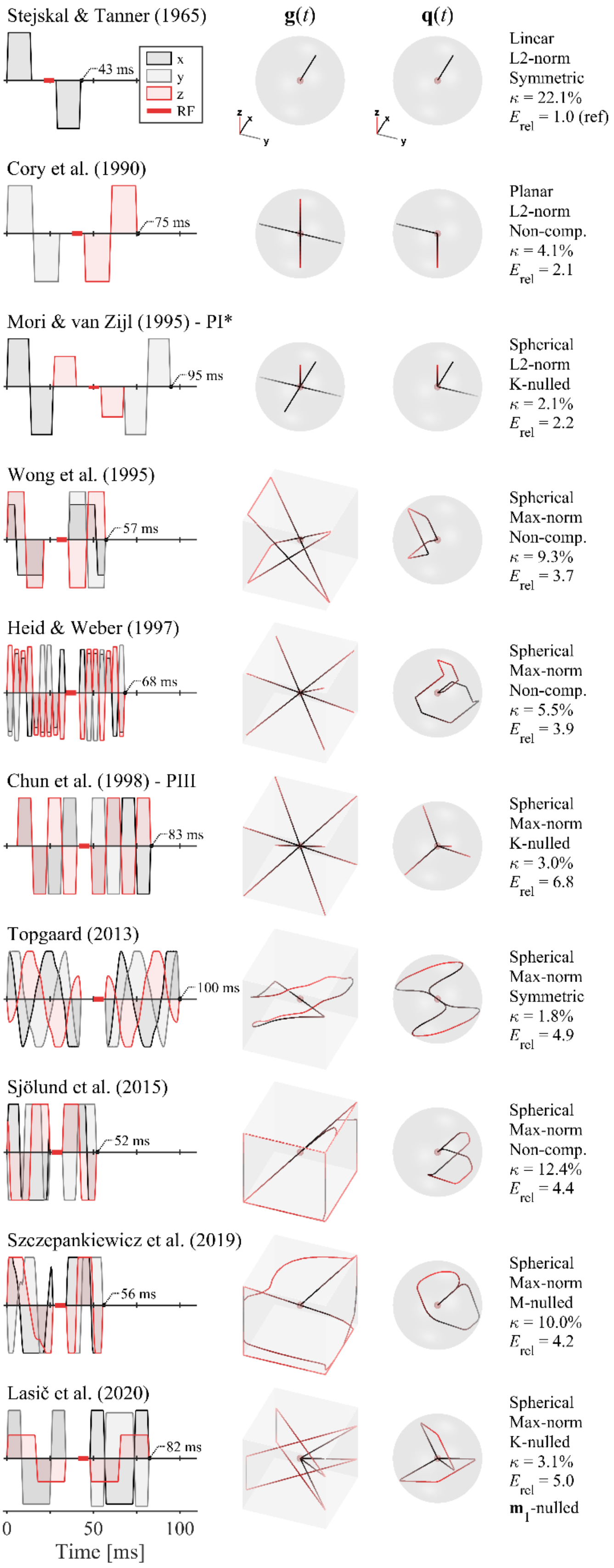

Figure 1 – A variety of gradient waveform designs and their dephasing vector trajectory in a spin-echo sequence. For comparability, all waveforms are adapted to yield $b$ = 2 ms/µm$^2$ at a minimal encoding time limiting the maximal gradient amplitude and slew rate to 80 mT/m and 100 T/m/s. We assume that the refocusing pulse lasts 8 ms and the encoding duration after the refocusing is 6 ms shorter than the duration before ($\delta_1 = \delta_2 + 6$ ms). Columns, from left to right, show the effective gradient waveform, the physical gradient trajectory, the dephasing q-vector trajectory, and a table of characteristics related to the b-tensor shape (section 2.1), optimization norm (section 3.1), concomitant gradient compensation (section 5.2), encoding efficiency ($\kappa$, section 2.3) and relative energy consumption ($E_{\text{rel}}$, section 3.2), as described throughout the paper. The monopolar pulsed field gradient by Stejskal and Tanner (1965) yields diffusion encoding along a single direction, also called linear b-tensor encoding (LTE). The waveform used by Cory et al. (1990) combines two orthogonal pairs of bipolar pulses to yield planar b-tensor encoding (PTE). This design allows arbitrarily directions for the two pairs and can therefore yield b-tensor shapes between LTE and PTE. Waveforms by Mori and van Zijl (1995), Wong et al. (1995), Heid and Weber (1997), Chun et al. (1998) and Topgaard (2013) were all designed to yield spherical b-tensor encoding (STE). The design by Lasic et al. (2020) can produce motion compensation of arbitrary order, and remaining designs by Sjölund et al. (2015), Szczepankiewicz et al. (2019d) generate optimal waveforms tailored to a given MRI system and sequence setup (only STE is shown). As indicated by the total duration next to the waveforms, the encoding efficiency varies widely (shorter times are more efficient). Note that we modified 'pattern I' in Mori and van Zijl (1995), denoted PI*, to improve its efficiency and yield compensation for concomitant gradient effects (K-nulling, section 5.2). The waveform by Wong et al. (1995) was split in two parts and placed around the refocusing pulse, according to the implementation in Butts et al. (1997). Finally, we note that the waveform by Heid and Weber (1997) is the basis for the 'one-scan-trace' design found at Siemens MRI systems (Dhital et al., 2018).

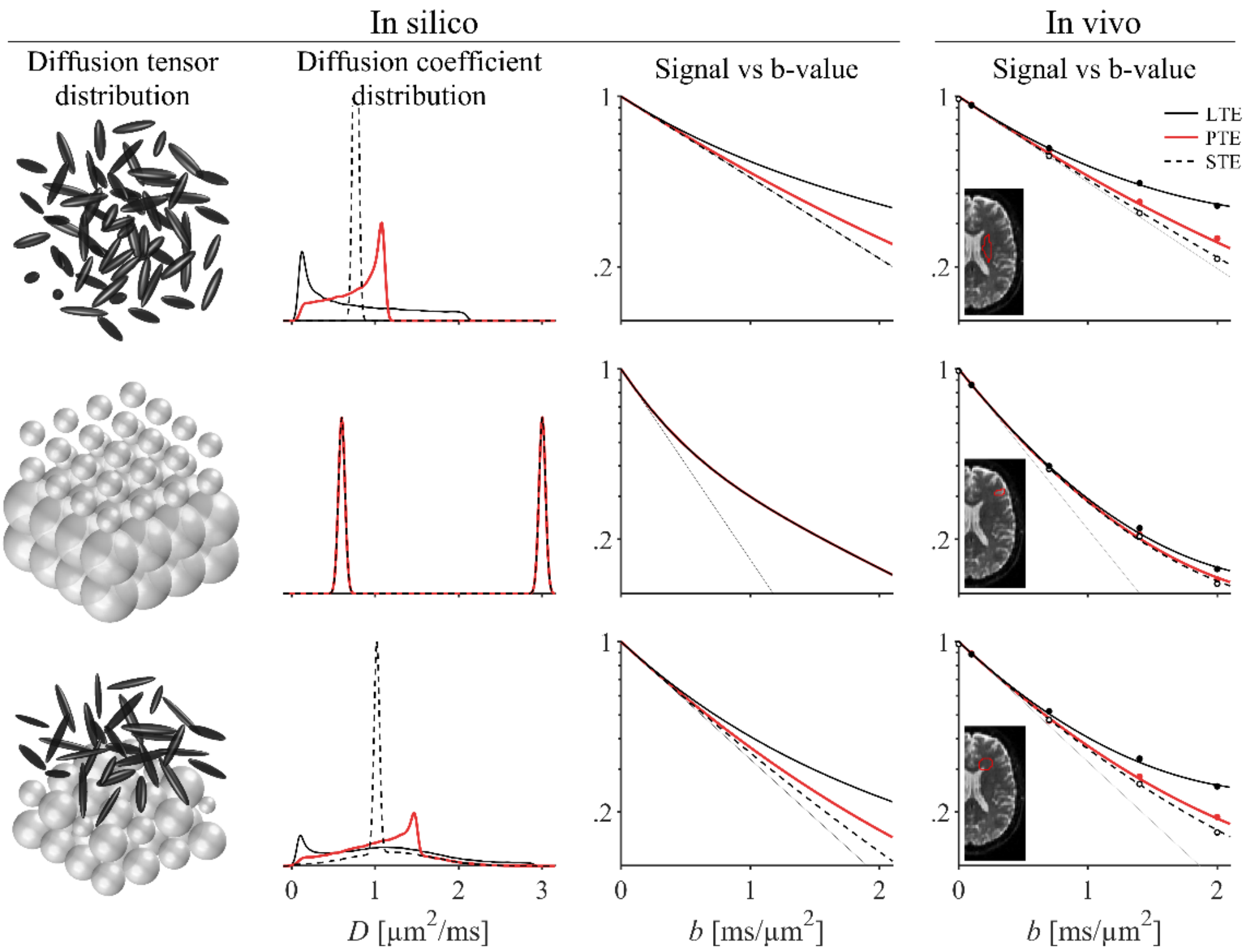

Figure 2 – The shape of the b-tensor influences the effect of diffusion anisotropy on the signal. The in silico examples show three diffusion tensor distributions, $P(\mathbf{D})$, with corresponding distributions of apparent diffusion coefficients, $P(D) = P((\mathbf{B}/b){:}\,\mathbf{D})$. From top to bottom they are randomly oriented anisotropic tensors; mixture of isotropic tensors with fast and slow diffusion; and mixture of anisotropic and isotropic diffusion tensors. The second column shows the effective distribution of apparent diffusion coefficients observed when using linear (LTE, solid black lines), planar (PTE, red lines) and spherical b-tensors (STE, broken black lines). The different distributions of diffusion coefficients manifest as different signal vs b-value curves (Eq. 9). For sufficiently large b-values, the signal is non-monoexponential in the presence of multiple diffusivities. Although this condition can be caused by markedly different diffusion tensor distributions, the three examples are indistinguishable if we can only make use of conventional diffusion encoding (Mitra, 1995, Szczepankiewicz et al., 2016). However, we may complement the measurement with b-tensors that have multiple shapes to isolate the contribution from microscopic diffusion anisotropy. This is the central motivation for using tensor-valued diffusion encoding. From a phenomenological perspective, the hallmark of 'microscopic diffusion anisotropy' is diverging signal between STE and all other b-tensor shapes, and the hallmark of 'heterogeneous isotropic diffusion' is non-monoexponential STE signal. The in vivo examples show similar signal behavior in three regions of healthy brain parenchyma, and it is the purpose of models and representations to infer the microstructure from the signal (Novikov et al., 2018a). The in vivo data used in this example is available online (Szczepankiewicz et al., 2019a).

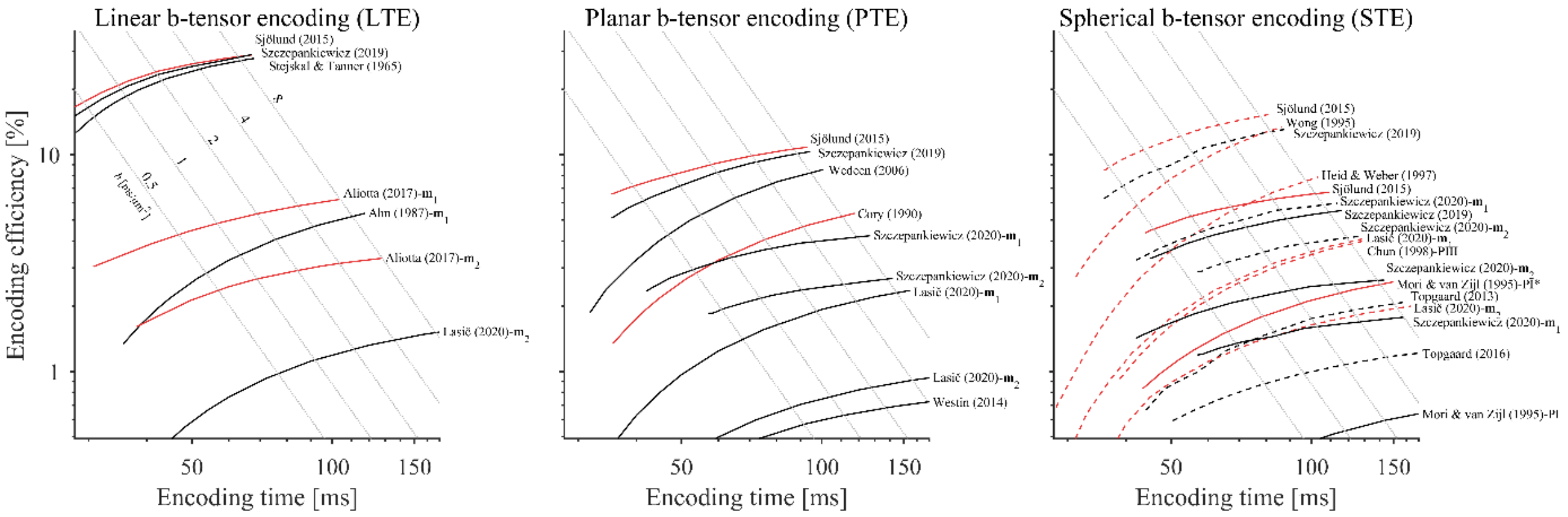

Figure 3 – Encoding efficiency ($\kappa$, Eq. 10) as a function of the total encoding time. The general trend is that linear encoding is the most efficient, followed by planar and spherical variants. Waveforms that are constrained by the max-norm (broken lines) have superior efficiency than those constrained by the L2-norm (solid lines, see section 3.1). However, we note that max-norm optimization does not allow arbitrary waveform rotation (Figure 4). Therefore, we show the max-norm only for spherical b-tensor encoding since they may not require any rotation. Furthermore, 'Maxwell-compensated' waveforms (black lines) have slightly lower efficiency compared to waveforms that may suffer errors from concomitant gradients (red lines) as described in section 5.2. For example, compare numerically optimized waveforms by Sjölund et al. (2015) and Szczepankiewicz et al. (2019d). The labels M1 and M2 indicate velocity and acceleration-compensation, respectively (section 4.3). Diagonal gray lines intersect the efficiency lies of each design at points where they yield b-values between 0.5 and 8 ms/µm$^2$. Although the optimal gradient waveform design is largely determined by the specifics of the application, the most versatile designs use the L2-norm and compensation for concomitant gradient effects (solid black lines). Finally, the impact of finite slew-rates can be seen as the amount of efficiency that is lost as the total encoding duration is reduced; waveforms that spend longer on slewing are the most affected (Sjölund et al., 2015).

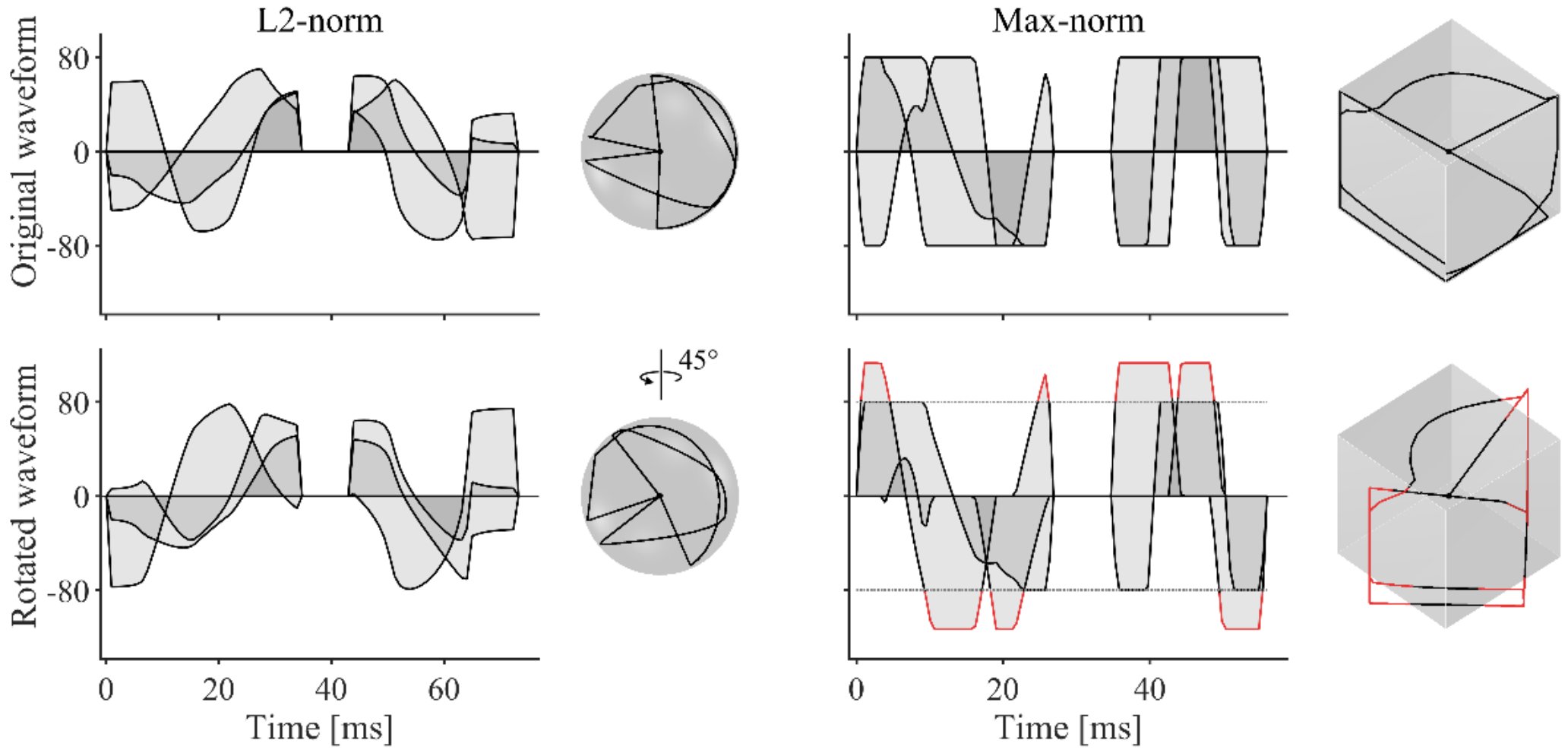

Figure 4 – Waveforms optimized with constraints on the L2-norm (Euclidean) and max-norm allow for different rotations. The gradient waveform on the left is inscribed within a sphere and can be freely rotated without violating the maximal gradient amplitude limitations. By contrast, the waveform on the right is inscribed within a cube and will therefore protrude through its surface when rotated. Naturally,

the max-norm still allows for rotations in steps of 90° around the x, y and z-axes, as well as axis permutations. In cases where the b-tensor does not have to be rotated, e.g., when we may assume that spherical b-tensor encoding is rotation invariant, the max-norm may provide a significant performance boost compared to the L2-norm. In this example, waveforms were numerically optimized (Sjölund et al., 2015, Szczepankiewicz et al., 2019d) to the conditions described in Figure 1, and yield $b$ = 2 ms/µm$^2$ in 73 ms and 56 ms, respectively. Naturally, any waveform that exceeds the capacity of the gradient amplitude or slew rate may be scaled down, or de-rated, to be within specifications, although this is at a cost to performance (Eq. 10).

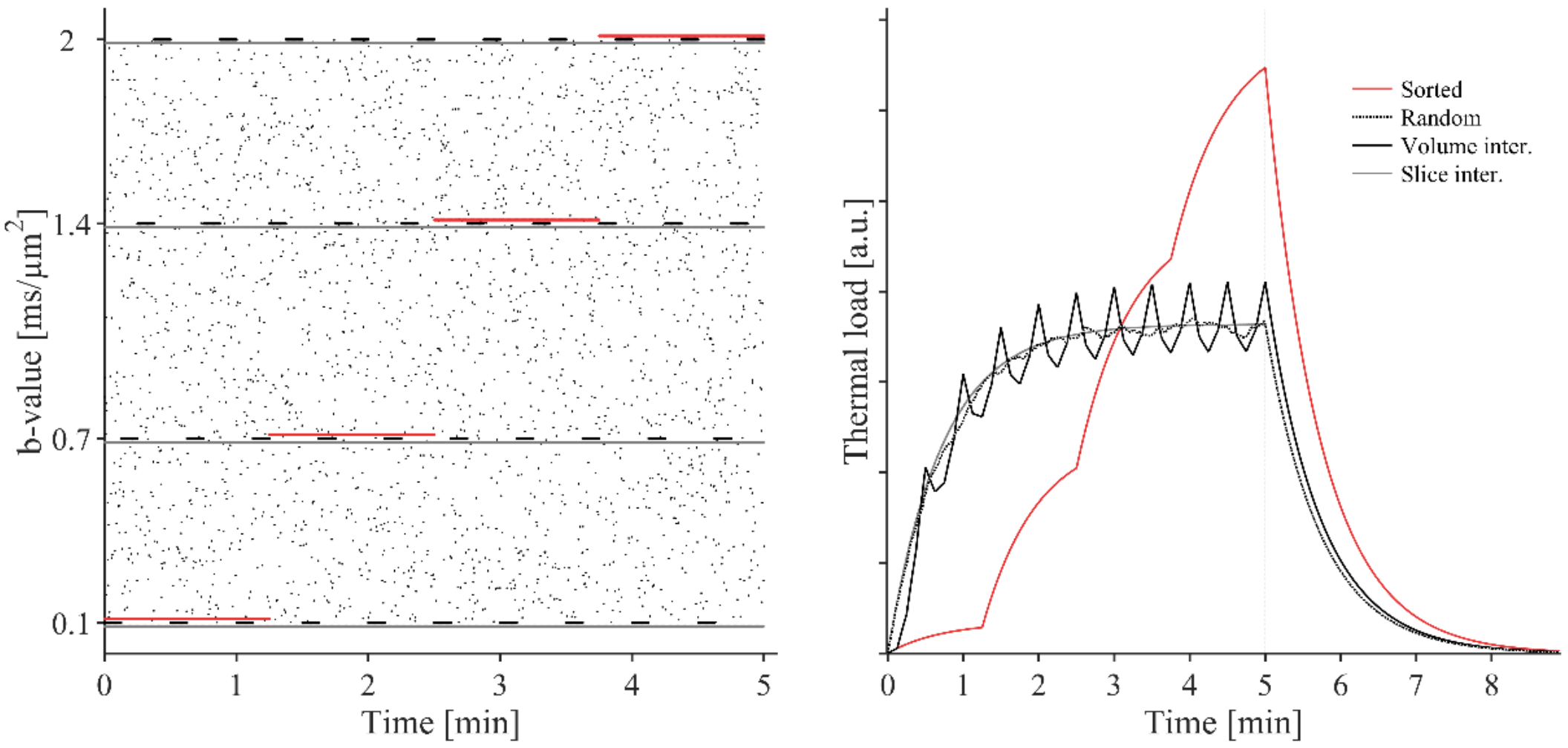

Figure 5 – The order of diffusion encoding determines the peak thermal load on the system. The sorted method uses four b-values executed in order, from low to high. The interleaved sampling schemes acquire the same b-values, but each consecutive volume or slice uses a different b-value. Similarly, the random sampling acquires consecutive slices with random diffusion weighting between 0.1 and 2.0 ms/µm$^2$. The figure shows that the peak thermal load is the largest when samples are acquired in order, i.e., executing the highest b-value many times in a row tends to heat the system in an unfavorable way compared to interleaved schemes (Hutter et al., 2018a). This example is based on Newton's law of cooling (Newton, 1701) for a hypothetical imaging setup that uses 60 slices per volume, 4 b-values, 10 repetitions per b-value, shots separated by 125 ms, and a system with a cooling time constant ln(2)/30 s$^{-1}$.

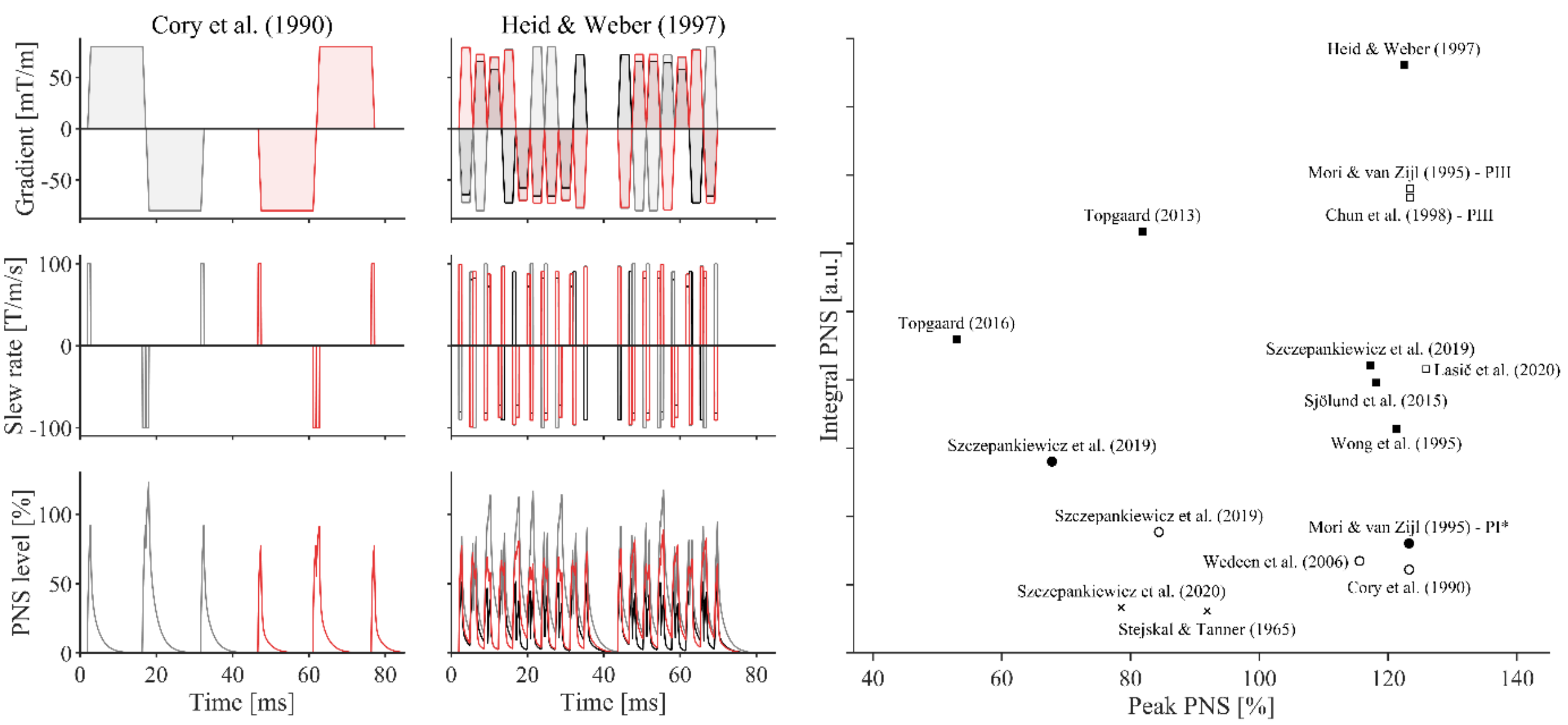

Figure 6 – Peripheral nerve stimulation (PNS) levels vary over time and are induced by rapid and sustained switching of the gradients. The plots on the left show waveforms with the highest peak and cumulative PNS among the tested waveforms, respectively. From the slew rate diagrams (second row), it is clear that high PNS occurs when gradients transition between maximal positive to negative gradients. We emphasize that all waveforms were limited to use the same maximal slew rate, yet they cover a wide range of PNS levels, as seen in the right plot. The PNS can be reduced by limiting the maximal slew rate for the whole waveform, or by limiting the slew rate in segments where switching is sustained for longer periods of time and/or in short succession. Moreover, the PNS response may depend on subject position or waveform rotation (Budinger et al., 1991, Lee et al., 2016). For example, the waveform on the left creates a larger peak PNS on the y-axis (gray) compared to the z-axis (red), even if the bipolar pairs are otherwise identical. In this figure, PNS values were calculated to visualize representative PNS response functions, the wide range of possible values, and the rotation variance using an in-house implementation (https://github.com/filip-szczepankiewicz/safe_pns_prediction) of the SAFE model (Hebrank and Gebhardt, 2000) assuming a representative MRI system.

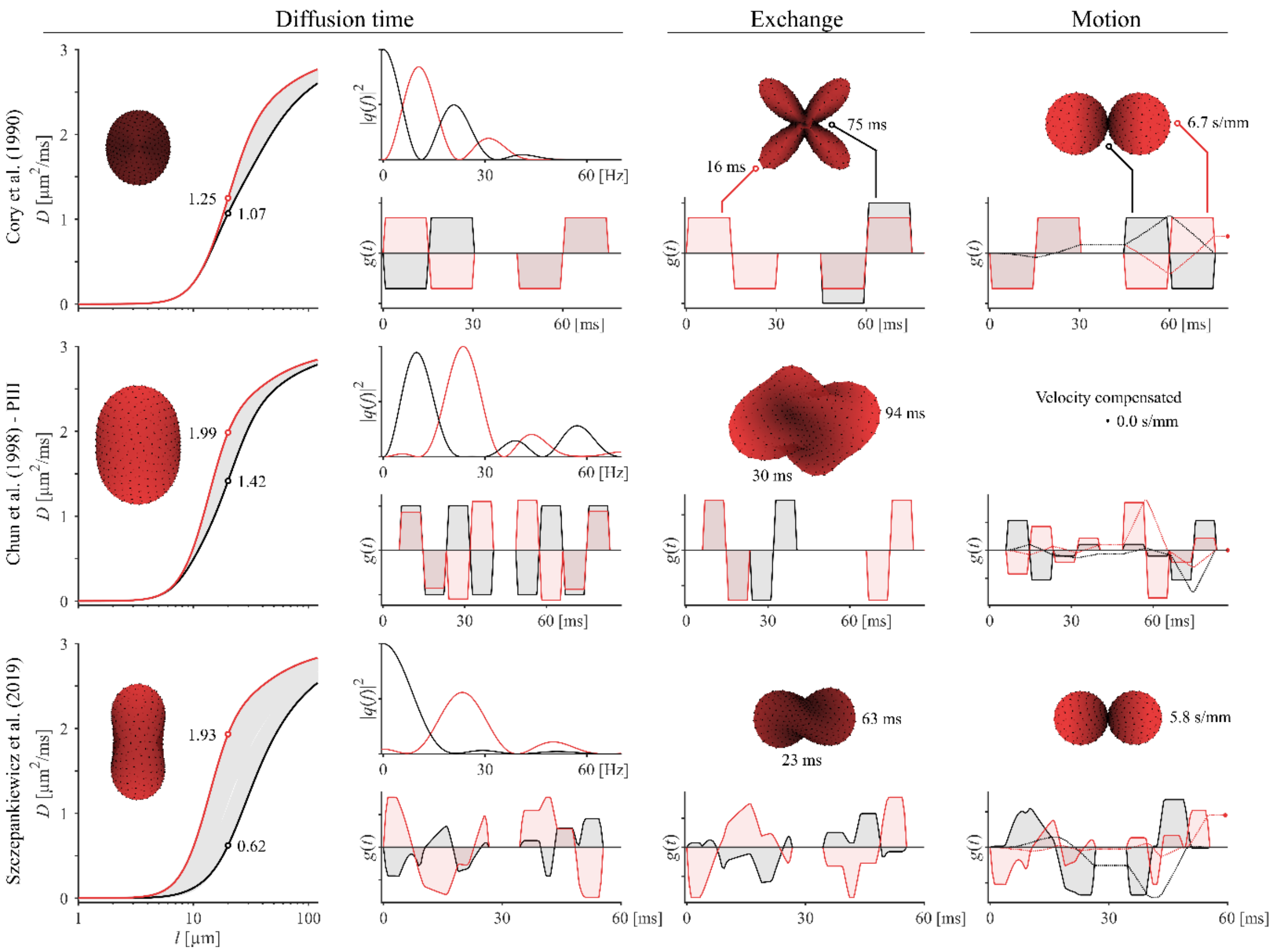

Figure 7 – The b-tensor describes encoding for Gaussian diffusion but most gradient waveforms also encode for features beyond the b-tensor. The first two columns visualize the effect of diffusion time where diffusion is restricted inside a one-dimensional stick with length $l$. The first column shows the apparent diffusion coefficients as a function of stick length and rotation; the inset glyph shows the apparent diffusivity for sticks with length $l$ = 20 µm directed along multiple directions. The second column shows the waveforms and encoding power spectra associated with the highest (red) and lowest (black) apparent diffusivity and encoding frequency. The trend is that faster gradient oscillations have more power at higher frequencies, equivalent to shorter diffusion times, and therefore detect a higher apparent diffusivity (Stepisnik, 1993). The third column shows a similar analysis of the exchange weighting time ($\Gamma$ in Eq. 24) (Ning et al., 2018). The glyphs show the exchange weighting along multiple directions, and the plotted waveforms are those that create the longest (red) and shortest (black) exchange times. In the final column, we show the vector-valued velocity encoding ($\mathbf{m}_1$ in Eq. 25) (Nalcioglu et al., 1986). Again, the sub-selected waveforms show maximal (red) and minimal (black) velocity encoding. For the selected example waveforms, we observe that diffusivity measured by DDE is relatively isotropic, even if the encoding power spectra are markedly different (similar $bV_\omega$ in Eq. 23). By contrast, the exchange weighting and velocity encoding are highly anisotropic. The waveform by Chun et al. (1998) has anisotropic diffusion time and exchange characteristics, but is velocity compensated in addition to its robustness to background gradient cross-terms (section 5.4). Finally, the design by Szczepankiewicz et al. (2019d) has a superior encoding efficiency and is compensated for concomitant gradients (section 5.2), but exhibits the highest spectral anisotropy.

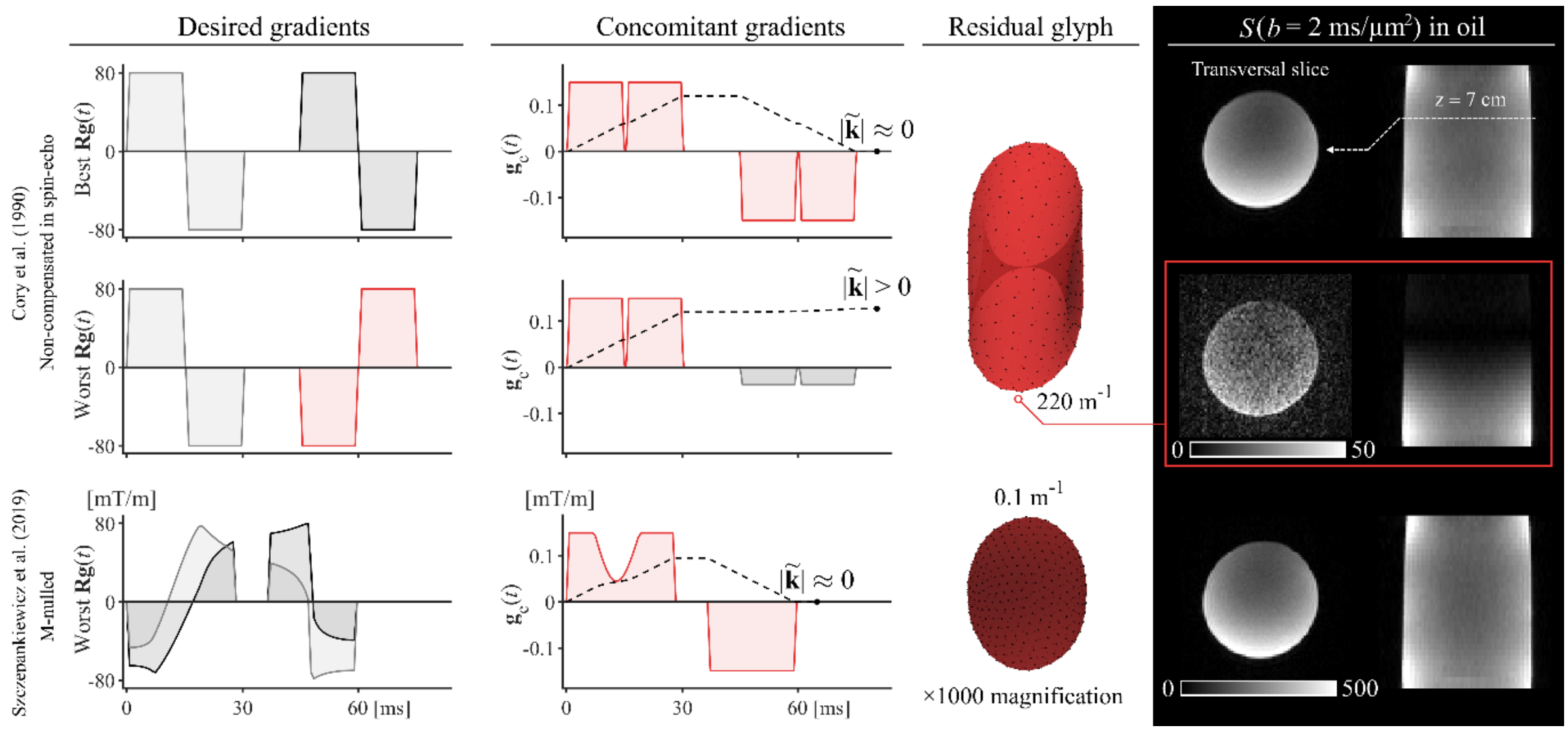

Figure 8 – The desired gradient waveform is always accompanied by concomitant gradients such that the actual gradient waveform is the sum of the two. Even if the desired gradient waveform is balanced, the concomitant waveform may not be, and their residual moment ($\mathbf{q}_c(\tau) = \tilde{\mathbf{k}}$) causes image artifacts that run the gamut between imperceptible to complete signal dropout. The columns, from left to right, show the desired gradient waveforms in different rotations ($\mathbf{R}\ \mathbf{g}(t)$), their concomitant gradient waveforms ($\mathbf{g}_c(t)$) and magnitude of the dephasing vector trajectory ($|\mathbf{q}_c(t)|$) at position $\mathbf{r}$ = [7 7 7] cm, glyphs of the residual dephasing vector $|\tilde{\mathbf{k}}|$ for an exhaustive set of waveform rotations, and signal maps in an oil phantom (Szczepankiewicz et al., 2019d). The first row shows double diffusion encoding where the gradient pulses are both in the x-y-plane. In this special case, the concomitant gradients cancel due to a symmetry in the contribution from the x and y axes (Eq. 29), and the dephasing vector at the end of encoding is negligible. By contrast, all other rotations result in a non-zero dephasing vector. For the worst rotation (middle row), this may result in a complete loss of signal, as seen 7 cm from the isocenter. In this oil phantom, the signal attenuation due to diffusion weighting is negligible, therefore, the severe signal attenuation in the middle row is an artifact caused by concomitant gradients. Note that this applies to the double diffusion encoding in a single spin-echo but can be avoided in double spin-echo sequences. Using an identical imaging setup, the worst rotation for a 'Maxwell-compensated' waveform has a negligible dephasing vector and no observable loss of signal (bottom row). Note that the glyphs show the residual dephasing for several orientations of the symmetry axis using the rotation method described in Szczepankiewicz et al. (2019a), where a final rotation around the symmetry axis is applied to find the worst case scenario. Details about the experiments and open source tools (https://github.com/markus-nilsson/md-dmri/tree/master/tools/cfa) for analysis of concomitant gradients and their effects can be found in Szczepankiewicz et al. (2019d).

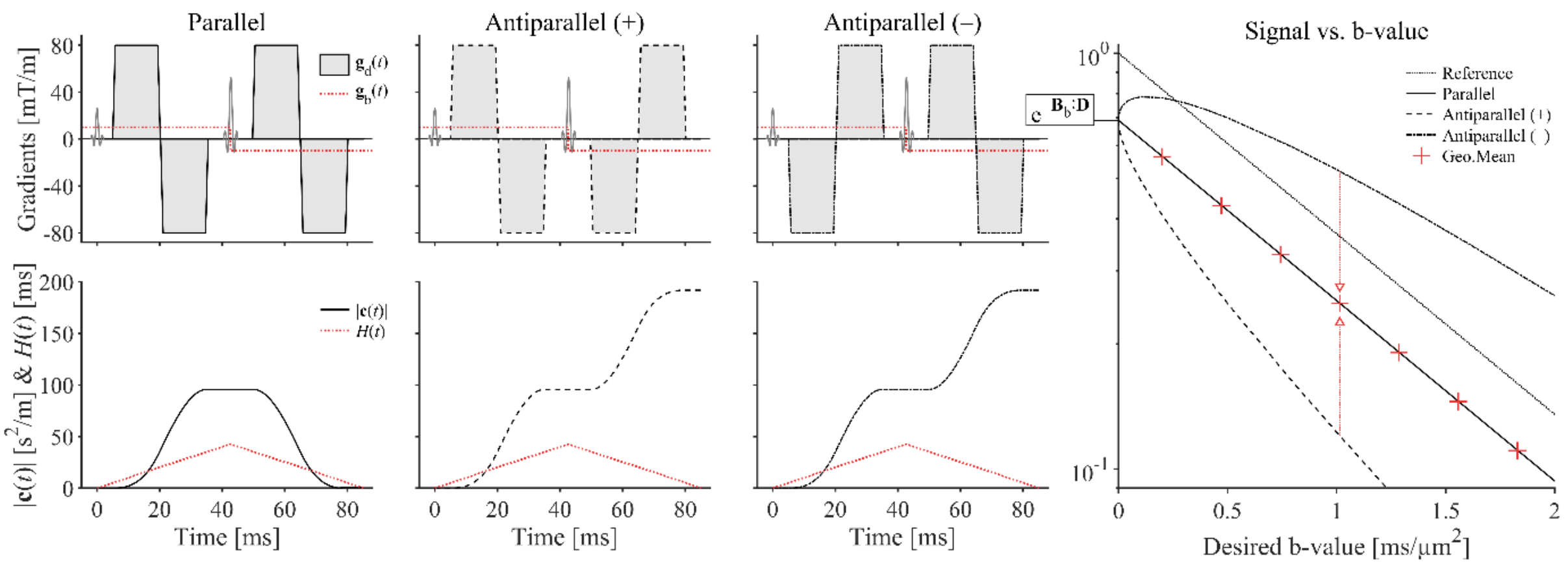

Figure 9 – The effect of background gradients will depend on their interaction with the desired gradient waveforms. The plots show three variants of double diffusion encoding waveforms in a spin-echo, and we introduce an exaggerated background gradient at 10 mT/m to visualize its effects. For simplicity, all gradient directions are assumed to be co-linear, and signal is simulated for isotropic diffusion. In the first column the desired gradient waveform creates parallel q-vectors whereas the remaining variants are antiparallel, using the nomenclature defined by Shemesh et al. (2016) rather than Hong and Dixon (1992). In the first case, the sign of the effective gradient waveform does not change before and after the refocusing such that multiplication with the reversed background gradient has the opposite effect. The waveform is also mirror-symmetric around the refocusing pulse, such that the time weighting factor ($H(t)$) has the same effect on both sides. The waveform is therefore cross-term-nulled ($\mathbf{c} = 0$, Eq. 34) and robust to cross-terms. The antiparallel variants exemplify the range of errors that can occur due to cross-terms. The 'antiparallel (+)' variant is initially directed along the same direction as the background gradient, causing a stronger diffusion encoding than expected, whereas the 'antiparallel (–)' variant starts out by opposing the background gradient, causing weaker encoding than expected. As such, the true b-value can be either higher or lower than the desired b-value. The effect of cross-terms and background gradient encoding on the diffusion weighted signal can be seen in the rightmost panel, where (+)/(–) variants causes under/over-estimation of signal, respectively. Note that the signal can exhibit hyper-attenuation, similar to the manifestation of a fast diffusivity or incoherent motion (Le Bihan, 1990), or signal that increases with desired b-value! These artifacts are not present for compensated gradient waveforms, or when the geometric average of signal is calculated from non-compensated variants (Neeman et al., 1991). Finally, we note that regardless of gradient waveform design, signal will be attenuated by the background gradient contribution to the b-tensor, i.e., signal is reduced by the factor $\exp(-\mathbf{B}_\mathrm{b}:\mathbf{D})$ which may be ignored in homogeneous substrates as shown in this example, but may cause additional errors in more complex substrates.

## References

AFZALI, M., TAX, C. M. W., CHATZIANTONIOU, C. & JONES, D. K. 2019. Comparison of Different Tensor Encoding Combinations in Microstructural Parameter Estimation. *2019 Ieee 16th International Symposium on Biomedical Imaging (Isbi 2019)*, 1471-1474.

AGGARWAL, M., JONES, M. V., CALABRESI, P. A., MORI, S. & ZHANG, J. 2012. Probing mouse brain microstructure using oscillating gradient diffusion MRI. *Magn Reson Med,* 67**,** 98-109.

AHLGREN, A., KNUTSSON, L., WIRESTAM, R., NILSSON, M., STÅHLBERG, F., TOPGAARD, D. & LASIČ, S. 2016. Quantification of microcirculatory parameters by joint analysis of flow-compensated and non-flow-compensated intravoxel incoherent motion (IVIM) data. *NMR Biomed,* 29**,** 640-9.

AHN, C. B., LEE, S. Y., NALCIOGLU, O. & CHO, Z. H. 1987. The effects of random directional distributed flow in nuclear magnetic resonance imaging. *Med Phys,* 14**,** 43-8.

ALEXANDER, A. L., TSURUDA, J. S. & PARKER, D. L. 1997. Elimination of Eddy Current Artifacts in Diffusion-Weighted Echo-Planar Images: The IJse of Bipolar Gradients. *Magnetic Resonance in Medicine,* 38(6)**,** 1016-21.

ALEXANDER, D. C. 2009. Modelling, Fitting and Sampling in Diffusion MRI.

ALIOTTA, E., MOULIN, K. & ENNIS, D. B. 2018. Eddy current-nulled convex optimized diffusion encoding (EN-CODE) for distortion-free diffusion tensor imaging with short echo times. *Magn Reson Med,* 79**,** 663-672.

ALIOTTA, E., WU, H. H. & ENNIS, D. B. 2017. Convex optimized diffusion encoding (CODE) gradient waveforms for minimum echo time and bulk motion-compensated diffusion-weighted MRI. *Magn Reson Med,* 77**,** 717-729.

ASSAF, Y., JOHANSEN-BERG, H. & THIEBAUT DE SCHOTTEN, M. 2019. The role of diffusion MRI in neuroscience. *NMR Biomed,* 32**,** e3762.

BAMMER, R., MARKL, M., BARNETT, A., ACAR, B., ALLEY, M. T., PELC, N. J., GLOVER, G. H. & MOSELEY, M. E. 2003. Analysis and generalized correction of the effect of spatial gradient field distortions in diffusion-weighted imaging. *Magn Reson Med,* 50**,** 560-9.

BARON, C. A., LEBEL, R. M., WILMAN, A. H. & BEAULIEU, C. 2012. The effect of concomitant gradient fields on diffusion tensor imaging. *Magn Reson Med,* 68**,** 1190-201.

BASSER, P. J., MATTIELLO, J. & LE BIHAN, D. 1994. MR diffusion tensor spectroscopy and imaging. *Biophys J,* 66**,** 259-67.

BATES, A. P., DADUCCI, A., SADEGHI, P. & CARUYER, E. 2020. A 4D Basis and Sampling Scheme for the Tensor Encoded Multi-Dimensional Diffusion MRI Signal. *IEEE Signal Processing Letters***,** 1-1.

BAUER, P., FERREIRA, J. A., SPANJJARD, S. & HOLLANDER, M. 2004. Innovative efficient gradient coil driver topology. *Apec 2004: Nineteenth Annual Ieee Applied Power Electronics Conference and Exposition, Vols 1-3***,** 1838-1843.

BEAULIEU, C. 2002. The basis of anisotropic water diffusion in the nervous system - a technical review. *NMR Biomed,* 15**,** 435-55.

BENJAMINI, D., KOMLOSH, M. E. & BASSER, P. J. 2017. Imaging Local Diffusive Dynamics Using Diffusion Exchange Spectroscopy MRI. *Phys Rev Lett,* 118**,** 158003.

BERNSTEIN, M. A., KING, K. F. & ZHOU, X. J. 2004. Part III, Gradients. *Handbook of MRI Pulse Sequences.* Elsevier Science & Technology. .

BERNSTEIN, M. A., ZHOU, X. J., POLZIN, J. A., KING, K. F., GANIN, A., PELC, N. J. & GLOVER, G. H. 1998. Concomitant gradient terms in phase contrast MR: analysis and correction. *Magn Reson Med,* 39**,** 300-8.

BLASCHE, M. 2017. Gradient Performance and Gradient Amplifier Power. *MAGNETOM Flash.* Siemens Healthcare.

BLÜMICH, B. 2016. k and q Dedicated to Paul Callaghan. *Journal of Magnetic Resonance*.

BREUER, F. A., BLAIMER, M., HEIDEMANN, R. M., MUELLER, M. F., GRISWOLD, M. A. & JAKOB, P. M. 2005. Controlled aliasing in parallel imaging results in higher acceleration (CAIPIRINHA) for multi-slice imaging. *Magn Reson Med,* 53**,** 684-91.

BUDDE, M. D. & SKINNER, N. P. 2018. Diffusion MRI in acute nervous system injury. *J Magn Reson,* 292**,** 137-148.

BUDINGER, T. F., FISCHER, H., HENTSCHEL, D., REINFELDER, H. E. & SCHMITT, F. 1991. Physiological effects of fast oscillating magnetic field gradients. *J Comput Assist Tomogr,* 15**,** 909-14.

BUTTS, K., DE CRESPIGNY, A., PAULY, J. M. & MOSELEY, M. 1996. Diffusion-weighted interleaved echo-planar imaging with a pair of orthogonal navigator echoes. *Magn Reson Med,* 35**,** 763-70.

BUTTS, K., PAULY, J., DE CRESPIGNY, A. & MOSELEY, M. 1997. Isotropic diffusion-weighted and spiral-navigated interleaved EPI for routine imaging of acute stroke. *Magn Reson Med,* 38**,** 741-9.

CALLAGHAN, P. T. 2011. Double wavevector encoding. *Translational Dynamics and Magnetic Resonance: Principles of Pulsed Gradient Spin Echo NMR.* Oxford University Press.

CALLAGHAN, P. T. & FURÓ, I. 2004. Diffusion-diffusion correlation and exchange as a signature for local order and dynamics. *The Journal of Chemical Physics*, 120**,** 4032-4038.

CALLAGHAN, P. T. & KOMLOSH, M. E. 2002. Locally anisotropic motion in a macroscopically isotropic system: displacement correlations measured using double pulsed gradient spin-echo NMR. *Magnetic Resonance in Chemistry,* 40**,** S15-S19.

CALLAGHAN, P. T. & STEPIŠNIK, J. 1996. Generalized Analysis of Motion Using Magnetic Field Gradients.

CAPRIHAN, A. & FUKUSHIMA, E. 1990. Flow Measurements by Nmr. *Physics Reports-Review Section of Physics Letters,* 198**,** 195-235.

CHEN, L., LIU, M., BAO, J., XIA, Y., ZHANG, J., ZHANG, L., HUANG, X. & WANG, J. 2013. The correlation between apparent diffusion coefficient and tumor cellularity in patients: a meta-analysis. *PLoS One,* 8**,** e79008.

CHUHUTIN, A., HANSEN, B. & JESPERSEN, S. N. 2017. Precision and accuracy of diffusion kurtosis estimation and the influence of b-value selection. *NMR Biomed,* 30.

CHUN, T., ULUG, A. M. & VAN ZIJL, P. C. 1998. Single-shot diffusion-weighted trace imaging on a clinical scanner. *Magn Reson Med,* 40**,** 622-8.

CLARK, C. A., HEDEHUS, M. & MOSELEY, M. E. 2001. Diffusion time dependence of the apparent diffusion tensor in healthy human brain and white matter disease. *Magn Reson Med,* 45**,** 1126-9.

COELHO, S., POZO, J. M., JESPERSEN, S. N. & FRANGI, A. F. 2019. Optimal experimental design for biophysical modelling in multidimensional diffusion MRI. *arXiv*.

COLVIN, D. C., YANKEELOV, T. E., DOES, M. D., YUE, Z., QUARLES, C. & GORE, J. C. 2008. New insights into tumor microstructure using temporal diffusion spectroscopy. *Cancer Res,* 68**,** 5941-7.

CORY, D. G., GARROWAY, A. N. & MILLER, J. B. 1990. Applications of Spin Transport as a Probe of Local Geometry. *Abstracts of Papers of the American Chemical Society,* 199**,** 105.

COTTAAR, M., SZCZEPANKIEWICZ, F., BASTIANI, M., HERNANDEZ-FERNANDEZ, M., SOTIROPOULOS, S. N., NILSSON, M. & JBABDI, S. 2020. Improved fibre dispersion estimation using b-tensor encoding. *Neuroimage,* 215**,** 116832.

DAVIDS, M., GUERIN, B., VOM ENDT, A., SCHAD, L. R. & WALD, L. L. 2019. Prediction of peripheral nerve stimulation thresholds of MRI gradient coils using coupled electromagnetic and neurodynamic simulations. *Magn Reson Med,* 81**,** 686-701.

DE ALMEIDA MARTINS, J. P., TAX, C. M. W., SZCZEPANKIEWICZ, F., JONES, D. K., WESTIN, C.-F. & TOPGAARD, D. 2020. Transferring principles of solid-state and Laplace NMR to the field of in vivo brain MRI. *Magnetic Resonance,* 1**,** 27-43.

DE GRAAF, R. A., BRAUN, K. P. & NICOLAY, K. 2001. Single-shot diffusion trace (1)H NMR spectroscopy. *Magn Reson Med,* 45**,** 741-8.

DE SWIET, T. M. & MITRA, P. P. 1996. Possible Systematic Errors in Single-Shot Measurements of the Trace of the Diffusion Tensor. *J Magn Reson B,* 111**,** 15-22.

DHITAL, B., KELLNER, E., KISELEV, V. G. & REISERT, M. 2018. The absence of restricted water pool in brain white matter. *Neuroimage,* 182**,** 398-406.

DHITAL, B., REISERT, M., KELLNER, E. & KISELEV, V. G. 2019. Intra-axonal diffusivity in brain white matter. *Neuroimage,* 189**,** 543-550.

DOES, M. D., PARSONS, E. C. & GORE, J. C. 2003. Oscillating gradient measurements of water diffusion in normal and globally ischemic rat brain. *Magn Reson Med,* 49**,** 206-15.

DORAN, S. J., CHARLES-EDWARDS, L., REINSBERG, S. A. & LEACH, M. O. 2005. A complete distortion correction for MR images: I. Gradient warp correction. *Phys Med Biol,* 50**,** 1343-61.

DROBNJAK, I. & ALEXANDER, D. C. 2011. Optimising time-varying gradient orientation for microstructure sensitivity in diffusion-weighted MR. *J Magn Reson,* 212**,** 344-54.

DU, Y. P., JOE ZHOU, X. & BERNSTEIN, M. A. 2002. Correction of concomitant magnetic field-induced image artifacts in nonaxial echo-planar imaging. *Magn Reson Med,* 48**,** 509-15.

ERIKSSON, S., ELBING, K., SODERMAN, O., LINDKVIST-PETERSSON, K., TOPGAARD, D. & LASIC, S. 2017. NMR quantification of diffusional exchange in cell suspensions with relaxation rate differences between intra and extracellular compartments. *PLoS One,* 12**,** e0177273.

ERIKSSON, S., LASIČ, S., NILSSON, M., WESTIN, C. F. & TOPGAARD, D. 2015. NMR diffusion-encoding with axial symmetry and variable anisotropy: Distinguishing between prolate and oblate microscopic diffusion tensors with unknown orientation distribution. *J Chem Phys,* 142**,** 104201.

ERIKSSON, S., LASIČ, S. & TOPGAARD, D. 2013. Isotropic diffusion weighting in PGSE NMR by magic-angle spinning of the q-vector. *J Magn Reson,* 226**,** 13-8.

FILLARD, P., DESCOTEAUX, M., GOH, A., GOUTTARD, S., JEURISSEN, B., MALCOLM, J., RAMIREZ-MANZANARES, A., REISERT, M., SAKAIE, K., TENSAOUTI, F., YO, T., MANGIN, J. F. & POUPON, C. 2011. Quantitative evaluation of 10 tractography algorithms on a realistic diffusion MR phantom. *Neuroimage,* 56**,** 220-34.

FINSTERBUSCH, J. 2008. Cross-term-compensated pulsed-gradient stimulated echo MR with asymmetric gradient pulse lengths. *J Magn Reson,* 193**,** 41-8.

FINSTERBUSCH, J. 2010. Double-spin-echo diffusion weighting with a modified eddy current adjustment. *Magn Reson Imaging,* 28**,** 434-40.

FINSTERBUSCH, J. 2011. Multiple-Wave-Vector Diffusion-Weighted NMR. *Annual Reports on NMR Spectroscopy,* 72**,** 225-299.

GALLICHAN, D., SCHOLZ, J., BARTSCH, A., BEHRENS, T. E., ROBSON, M. D. & MILLER, K. L. 2010. Addressing a systematic vibration artifact in diffusion-weighted MRI. *Hum Brain Mapp,* 31**,** 193-202.

GORE, J. C., XU, J., COLVIN, D. C., YANKEELOV, T. E., PARSONS, E. C. & DOES, M. D. 2010. Characterization of tissue structure at varying length scales using temporal diffusion spectroscopy. *NMR Biomed,* 23**,** 745-56.

GREBENKOV, D. S. 2007. NMR survey of reflected Brownian motion. *Reviews of Modern Physics,* 79**,** 1077-1137.

HAACKE, E. M. & LENZ, G. W. 1987. Improving Mr Image Quality in the Presence of Motion by Using Rephasing Gradients. *American Journal of Roentgenology,* 148**,** 1251-1258.

HABIB, J., AUER, D. P. & MORGAN, P. S. 2010. A quantitative analysis of the benefits of cardiac gating in practical diffusion tensor imaging of the brain. *Magn Reson Med,* 63**,** 1098-103.

HAHN, E. L. 1950. Spin Echoes. *Physical Review,* 80**,** 580-594.

HAHN, E. L. 1960. Detection of sea-water motion by nuclear precession. *Journal of Geophysical Research,* 65**,** 776-777.

HAM, C. L., ENGELS, J. M., VAN DE WIEL, G. T. & MACHIELSEN, A. 1997. Peripheral nerve stimulation during MRI: effects of high gradient amplitudes and switching rates. *J Magn Reson Imaging,* 7**,** 933-7.

HARGREAVES, B. A., NISHIMURA, D. G. & CONOLLY, S. M. 2004. Time-optimal multidimensional gradient waveform design for rapid imaging. *Magn Reson Med,* 51**,** 81-92.

HEBRANK, F. X. & GEBHARDT, M. SAFE-Model - A New Method for Predicting Peripheral Nerve Stimulations in MRI. Proc. Intl. Soc. Mag. Reson. Med. 8, 2000 Denver, CO, USA. 2007.

HEDEEN, R. A. & EDELSTEIN, W. A. 1997. Characterization and prediction of gradient acoustic noise in MR imagers. *Magn Reson Med,* 37**,** 7-10.

HEID, O. & WEBER, J. Diffusion Tensor Trace Pulse Sequences. Proc. Intl. Soc. Magn. Reson. Med. 5, 1997 Vancouver, Canada.

HEISMANN, B., OTT, M. & GRODZKI, D. 2015. Sequence-based acoustic noise reduction of clinical MRI scans. *Magn Reson Med,* 73**,** 1104-9.

HENRIQUES, R. N., JESPERSEN, S. N. & SHEMESH, N. 2019. Microscopic anisotropy misestimation in spherical-mean single diffusion encoding MRI. *Magn Reson Med*.

HENRIQUES, R. N., JESPERSEN, S. N. & SHEMESH, N. 2020. Correlation tensor magnetic resonance imaging. *Neuroimage,* 211**,** 116605.

HIDALGO-TOBON, S. S. 2010. Theory of gradient coil design methods for magnetic resonance imaging. *Concepts in Magnetic Resonance Part A,* 36A**,** 223-242.

HILTUNEN, J., HARI, R., JOUSMAKI, V., MULLER, K., SEPPONEN, R. & JOENSUU, R. 2006. Quantification of mechanical vibration during diffusion tensor imaging at 3 T. *Neuroimage,* 32**,** 93-103.

HONG, X. & DIXON, T. W. 1992. Measuring diffusion in inhomogeneous systems in imaging mode using antisymmetric sensitizing gradients. *Journal of Magnetic Resonance (1969),* 99**,** 561-570.

HORSFIELD, M. A. & JONES, D. K. 2002. Applications of diffusion-weighted and diffusion tensor MRI to white matter diseases - a review. *NMR Biomed,* 15**,** 570-7.

HUTTER, J., NILSSON, M., CHRISTIAENS, D., SCHNEIDER, T., PRICE, A. N., HAJNAL, J. V. & SZCZEPANKIEWICZ, F. Highly efficient diffusion MRI by slice-interleaved free-waveform imaging (SIFI). Proc. Intl. Soc. Mag. Reson. Med. 26, 2018a Paris, France.

HUTTER, J., PRICE, A. N., CORDERO-GRANDE, L., MALIK, S., FERRAZZI, G., GASPAR, A., HUGHES, E. J., CHRISTIAENS, D., MCCABE, L., SCHNEIDER, T., RUTHERFORD, M. A. & HAJNAL, J. V. 2018b. Quiet echo planar imaging for functional and diffusion MRI. *Magn Reson Med,* 79**,** 1447-1459.

HUTTER, J., TOURNIER, J. D., PRICE, A. N., CORDERO-GRANDE, L., HUGHES, E. J., MALIK, S., STEINWEG, J., BASTIANI, M., SOTIROPOULOS, S. N., JBABDI, S., ANDERSSON, J., EDWARDS, A. D. & HAJNAL, J. V. 2018c. Time-efficient and flexible design of optimized multishell HARDI diffusion. *Magn Reson Med,* 79**,** 1276-1292.

IANUS, A., DROBNJAK, I. & ALEXANDER, D. C. 2016. Model-based estimation of microscopic anisotropy using diffusion MRI: a simulation study. *NMR Biomed,* 29**,** 672-85.

IANUS, A., JESPERSEN, S. N., SERRADAS DUARTE, T., ALEXANDER, D. C., DROBNJAK, I. & SHEMESH, N. 2018. Accurate estimation of microscopic diffusion anisotropy and its time dependence in the mouse brain. *Neuroimage,* 183**,** 934-949.

IANUS, A. & SHEMESH, N. 2017. Incomplete initial nutation diffusion imaging: An ultrafast, single-scan approach for diffusion mapping. *Magn Reson Med*.

IRELAND, C. M., GIAQUINTO, R. O., LOEW, W., TKACH, J. A., PRATT, R. G., KLINE-FATH, B. M., MERHAR, S. L. & DUMOULIN, C. L. 2015. A novel acoustically quiet coil for neonatal MRI system. *Concepts Magn Reson Part B Magn Reson Eng,* 45**,** 107-114.

IRFANOGLU, M. O., SARLLS, J., NAYAK, A. & PIERPAOLI, C. 2019. Evaluating corrections for Eddy-currents and other EPI distortions in diffusion MRI: methodology and a dataset for benchmarking. *Magn Reson Med,* 81**,** 2774-2787.

IRFANOGLU, M. O., WALKER, L., SARLLS, J., MARENCO, S. & PIERPAOLI, C. 2012. Effects of image distortions originating from susceptibility variations and concomitant fields on diffusion MRI tractography results. *Neuroimage,* 61**,** 275-88.

IRNICH, W. & SCHMITT, F. 1995. Magnetostimulation in MRI. *Magn Reson Med,* 33**,** 619-23.

IVANOV, E. N., POGROMSKY, A. Y., VAN DEN BRINK, J. S. & ROODA, J. E. 2010. Optimization of duty cycles for MRI scanners. *Concepts in Magnetic Resonance Part B: Magnetic Resonance Engineering,* 37B**,** 180-192.

JARA, H. & WEHRLI, F. W. 1994. Determination of background gradients with diffusion MR imaging. *J Magn Reson Imaging,* 4**,** 787-97.

JELESCU, I. O. & BUDDE, M. D. 2017. Design and Validation of Diffusion MRI Models of White Matter. *Frontiers in Physics,* 5.

JELESCU, I. O., PALOMBO, M., BAGNATO, F. & SCHILLING, K. G. 2020. Challenges for biophysical modeling of microstructure. *J Neurosci Methods,* 344**,** 108861.

JELLISON, B. J., FIELD, A. S., MEDOW, J., LAZAR, M., SALAMAT, M. S. & ALEXANDER, A. L. 2004. Diffusion tensor imaging of cerebral white matter: a pictorial review of physics, fiber tract anatomy, and tumor imaging patterns. *AJNR Am J Neuroradiol,* 25**,** 356-69.

JENSEN, J. H., HUI, E. S. & HELPERN, J. A. 2014. Double-pulsed diffusional kurtosis imaging. *NMR Biomed,* 27**,** 363-370.

JESPERSEN, S. N., LUNDELL, H., SØNDERBY, C. K. & DYRBY, T. B. 2013. Orientationally invariant metrics of apparent compartment eccentricity from double pulsed field gradient diffusion experiments. *NMR Biomed,* 26**,** 1647-62.

JESPERSEN, S. N., OLESEN, J. L., IANUS, A. & SHEMESH, N. 2019. Effects of nongaussian diffusion on "isotropic diffusion" measurements: An ex-vivo microimaging and simulation study. *J Magn Reson,* 300**,** 84-94.

JEZZARD, P. & BALABAN, R. S. 1995. Correction for geometric distortion in echo planar images from B0 field variations. *Magn Reson Med,* 34**,** 65-73.

JEZZARD, P., BARNETT, A. S. & PIERPAOLI, C. 1998. Characterization of and correction for eddy current artifacts in echo planar diffusion imaging. *Magn Reson Med,* 39**,** 801-12.

JIAN, B., VEMURI, B. C., OZARSLAN, E., CARNEY, P. R. & MARECI, T. H. 2007. A novel tensor distribution model for the diffusion-weighted MR signal. *Neuroimage,* 37**,** 164-76.

JONES, D. K. 2008. Studying connections in the living human brain with diffusion MRI. *Cortex,* 44**,** 936-52.

JONES, D. K., ALEXANDER, D. C., BOWTELL, R., CERCIGNANI, M., DELL'ACQUA, F., MCHUGH, D. J., MILLER, K. L., PALOMBO, M., PARKER, G. J. M., RUDRAPATNA, U. S. & TAX, C. M. W. 2018. Microstructural imaging of the human brain with a 'super-scanner': 10 key advantages of ultra-strong gradients for diffusion MRI. *Neuroimage,* 182**,** 8-38.

JONES, D. K. & BASSER, P. J. 2004. "Squashing peanuts and smashing pumpkins": how noise distorts diffusion-weighted MR data. *Magn Reson Med,* 52**,** 979-93.

JOVICICH, J., CZANNER, S., GREVE, D., HALEY, E., VAN DER KOUWE, A., GOLLUB, R., KENNEDY, D., SCHMITT, F., BROWN, G., MACFALL, J., FISCHL, B. & DALE, A. 2006. Reliability in multi-site structural MRI studies: effects of gradient non-linearity correction on phantom and human data. *Neuroimage,* 30**,** 436-43.

KARLICEK, R. F. & LOWE, I. J. 1980. A modified pulsed gradient technique for measuring diffusion in the presence of large background gradients. *Journal of Magnetic Resonance (1969),* 37**,** 75-91.

KINGSLEY, P. B. 2006. Introduction to diffusion tensor imaging mathematics: Part II. Anisotropy, diffusion-weighting factors, and gradient encoding schemes. *Concepts in Magnetic Resonance Part A,* 28A**,** 123-154.

KOMLOSH, M. E., HORKAY, F., FREIDLIN, R. Z., NEVO, U., ASSAF, Y. & BASSER, P. J. 2007. Detection of microscopic anisotropy in gray matter and in a novel tissue phantom using double Pulsed Gradient Spin Echo MR. *J Magn Reson,* 189**,** 38-45.

KÄRGER, J. 1985. Nmr Self-Diffusion Studies in Heterogeneous Systems. *Advances in Colloid and Interface Science,* 23**,** 129-148.

LAMPINEN, B., SZCZEPANKIEWICZ, F., MARTENSSON, J., VAN WESTEN, D., HANSSON, O., WESTIN, C. F. & NILSSON, M. 2020a. Towards unconstrained compartment modeling in white matter using diffusion-relaxation MRI with tensor-valued diffusion encoding. *Magn Reson Med*.

LAMPINEN, B., SZCZEPANKIEWICZ, F., NOVEN, M., VAN WESTEN, D., HANSSON, O., ENGLUND, E., MARTENSSON, J., WESTIN, C. F. & NILSSON, M. 2019. Searching for the neurite density with diffusion MRI: Challenges for biophysical modeling. *Hum Brain Mapp,* 40**,** 2529-2545.

LAMPINEN, B., SZCZEPANKIEWICZ, F., VAN WESTEN, D., ENGLUND, E., SUNDGREN, P., LÄTT, J., STÅHLBERG, F. & NILSSON, M. 2016. Optimal experimental design for filter exchange imaging: Apparent exchange rate measurements in the healthy brain and in intracranial tumors. *Magn Reson Med*.

LAMPINEN, B., ZAMPELI, A., BJORKMAN-BURTSCHER, I. M., SZCZEPANKIEWICZ, F., KALLEN, K., COMPAGNO STRANDBERG, M. & NILSSON, M. 2020b. Tensor-valued diffusion MRI differentiates cortex and white matter in malformations of cortical development associated with epilepsy. *Epilepsia*.

LARKMAN, D. J., HAJNAL, J. V., HERLIHY, A. H., COUTTS, G. A., YOUNG, I. R. & EHNHOLM, G. 2001. Use of multicoil arrays for separation of signal from multiple slices simultaneously excited. *Journal of Magnetic Resonance Imaging,* 13**,** 313-317.

LASIČ, S., LUNDELL, H., SZCZEPANKIEWICZ, F., NILSSON, M., SCHNEIDER, J. E. & TEH, I. Time-dependent and anisotropic diffusion in the heart: linear and spherical tensor encoding with varying degree of motion compensation. Proc. Intl. Soc. Magn. Reson. Med. 28, 2020 Sydney, Australia.

LASIČ, S., NILSSON, M., LATT, J., STAHLBERG, F. & TOPGAARD, D. 2011. Apparent exchange rate mapping with diffusion MRI. *Magn Reson Med,* 66**,** 356-65.

LASIC, S., SZCZEPANKIEWICZ, F., DALL'ARMELLINA, E., DAS, A., KELLY, C., PLEIN, S., SCHNEIDER, J. E., NILSSON, M. & TEH, I. 2020. Motion-compensated b-tensor encoding for in vivo cardiac diffusion-weighted imaging. *NMR Biomed,* 33**,** e4213.

LASIČ, S., SZCZEPANKIEWICZ, F., ERIKSSON, S., NILSSON, M. & TOPGAARD, D. 2014. Microanisotropy imaging: quantification of microscopic diffusion anisotropy and orientational order parameter by diffusion MRI with magic-angle spinning of the q-vector. *Frontiers in Physics,* 2**,** 11.

LAUN, F. B. & KUDER, T. A. 2013. Diffusion pore imaging with generalized temporal gradient profiles. *Magn Reson Imaging,* 31**,** 1236-44.

LAWRENZ, M. & FINSTERBUSCH, J. 2019. Detection of microscopic diffusion anisotropy in human cortical gray matter in vivo with double diffusion encoding. *Magn Reson Med,* 81**,** 1296-1306.

LAWRENZ, M., KOCH, M. A. & FINSTERBUSCH, J. 2010. A tensor model and measures of microscopic anisotropy for double-wave-vector diffusion-weighting experiments with long mixing times. *J Magn Reson,* 202**,** 43-56.

LE BIHAN, D. 1990. Magnetic resonance imaging of perfusion. *Magn Reson Med,* 14**,** 283-92.

LE BIHAN, D. 2013. Apparent diffusion coefficient and beyond: what diffusion MR imaging can tell us about tissue structure. *Radiology,* 268**,** 318-22.

LE BIHAN, D., BRETON, E., LALLEMAND, D., GRENIER, P., CABANIS, E. & LAVAL-JEANTET, M. 1986. MR imaging of intravoxel incoherent motions: application to diffusion and perfusion in neurologic disorders. *Radiology,* 161**,** 401-7.

LEBEL, C., TREIT, S. & BEAULIEU, C. 2019. A review of diffusion MRI of typical white matter development from early childhood to young adulthood. *NMR Biomed,* 32**,** e3778.

LEE, S. K., MATHIEU, J. B., GRAZIANI, D., PIEL, J., BUDESHEIM, E., FIVELAND, E., HARDY, C. J., TAN, E. T., AMM, B., FOO, T. K., BERNSTEIN, M. A., HUSTON, J., 3RD, SHU, Y. & SCHENCK, J. F. 2016. Peripheral nerve stimulation characteristics of an asymmetric head-only gradient coil compatible with a high-channel-count receiver array. *Magn Reson Med,* 76**,** 1939-1950.

LEMBERSKIY, G., ROSENKRANTZ, A. B., VERAART, J., TANEJA, S. S., NOVIKOV, D. S. & FIEREMANS, E. 2017. Time-Dependent Diffusion in Prostate Cancer. *Invest Radiol,* 52**,** 405-411.

LIAN, J., WILIAMS, D. S. & LOWE, I. J. 1994. Magnetic resonance imaging of diffusion in the presence of background gradients and imaging of background gradients. *Journal of magnetic Resonance,* Series A 106**,** 65-74.

LUNDELL, H. & LASIČ, S. 2020. Diffusion Encoding with General Gradient Waveforms. *In:* TOPGAARD, D. (ed.) *Advanced Diffusion Encoding Methods in MRI.* Royal Society of Chemistry, 2020.

LUNDELL, H., LASIČ, S., SZCZEPANKIEWICZ, F., NILSSON, M., TOPGAARD, D., SCHNEIDER, J. E. & TEH, I. Stay on the beat: tuning in on time-dependent diffusion in the heart. Proc. Intl. Soc. Magn. Reson. Med. 28, 2020 Virtual.

LUNDELL, H., NILSSON, M., DYRBY, T. B., PARKER, G. J. M., CRISTINACCE, P. L. H., ZHOU, F. L., TOPGAARD, D. & LASIČ, S. 2019. Multidimensional diffusion MRI with spectrally modulated gradients reveals unprecedented microstructural detail. *Sci Rep,* 9**,** 9026.

LUNDELL, H., NILSSON, M., WESTIN, C.-F., TOPGAARD, D. & LASIČ, S. Spectral anisotropy in multidimensional diffusion encoding. Proc. Intl. Soc. Magn. Reson. Med. 26, 2018 Paris, France.

MANSFIELD, P. 1977. Multi-Planar Image-Formation Using Nmr Spin Echoes. *Journal of Physics C-Solid State Physics,* 10**,** L55-L58.

MANSFIELD, P., CHAPMAN, B. L., BOWTELL, R., GLOVER, P., COXON, R. & HARVEY, P. R. 1995. Active acoustic screening: reduction of noise in gradient coils by Lorentz force balancing. *Magn Reson Med,* 33**,** 276-81.

MANSFIELD, P., GLOVER, P. & BOWTELL, R. 1994. Active acoustic screening: design principles for quiet gradient coils in MRI. *Meas Sci. Techno.,* 5**,** 1021-1025.

MANSFIELD, P., GLOVER, P. M. & BEAUMONT, J. 1998. Sound generation in gradient coil structures for MRI. *Magn Reson Med,* 39**,** 539-50.

MANSFIELD, P. & HARVEY, P. R. 1993. Limits to neural stimulation in echo-planar imaging. *Magn Reson Med,* 29**,** 746-58.

MATTIELLO, J., BASSER, P. J. & LEBIHAN, D. 1997. The b matrix in diffusion tensor echo-planar imaging. *Magnetic Resonance in Medicine,* 37**,** 292-300.

MCJURY, M. & SHELLOCK, F. G. 2000. Auditory Noise Associated With MR Procedures: A Review. *JOURNAL OF MAGNETIC RESONANCE IMAGING,* 12**,** 37-45.

MEIER, C., ZWANGER, M., FEIWEIER, T. & PORTER, D. 2008. Concomitant field terms for asymmetric gradient coils: consequences for diffusion, flow, and echo-planar imaging. *Magn Reson Med,* 60**,** 128-34.

MENDITTON, A., PATRIARCA, M. & MAGNUSSON, B. 2006. Understanding the meaning of accuracy. *Accred Qual Assur*.

MESRI, H. Y., DAVID, S., VIERGEVER, M. A. & LEEMANS, A. 2019. The adverse effect of gradient nonlinearities on diffusion MRI: From voxels to group studies. *Neuroimage,* 205**,** 116127.

MITRA, P. 1995. Multiple wave-vector extensions of the NMR pulsed-field-gradient spin-echo diffusion measurement. *Physical Review B,* 51**,** 15074-15078.

MITRA, P. P., SEN, P. N. & SCHWARTZ, L. M. 1993. Short-time behavior of the diffusion coefficient as a geometrical probe of porous media. *Phys Rev B Condens Matter,* 47**,** 8565-8574.

MOFFAT, B. A., HALL, D. E., STOJANOVSKA, J., MCCONVILLE, P. J., MOODY, J. B., CHENEVERT, T. L., REHEMTULLA, A. & ROSS, B. D. 2004. Diffusion imaging for evaluation of tumor therapies in preclinical animal models. *MAGMA,* 17**,** 249-59.

MORAN, P. R. 1982. A flow velocity zeugmatographic interlace for NMR imaging in humans. *Magn Reson Imaging,* 1**,** 197-203.

MORI, S. & VAN ZIJL, P. 1995. Diffusion Weighting by the Trace of the Diffusion Tensor within a Single Scan. *Magn Reson Med,* 33**,** 41-52.

MOSELEY, M. E., COHEN, Y., MINTOROVITCH, J., CHILEUITT, L., SHIMIZU, H., KUCHARCZYK, J., WENDLAND, M. F. & WEINSTEIN, P. R. 1990a. Early detection of regional cerebral ischemia in cats: comparison of diffusion- and T2-weighted MRI and spectroscopy. *Magn Reson Med,* 14**,** 330-46.

MOSELEY, M. E., KUCHARCZYK, J., MINTOROVITCH, J., COHEN, Y., KURHANEWICZ, J., DERUGIN, N., ASGARI, H. & NORMAN, D. 1990b. Diffusion-weighted MR imaging of acute stroke: correlation with T2-weighted and magnetic susceptibility-enhanced MR imaging in cats. *AJNR Am J Neuroradiol,* 11**,** 423-9.

NAJAC, C., LUNDELL, H., BULK, M., KAN, H. E., WEBB, A. G. & RONEN, I. Estimating compartment- and cell-specific microscopic anisotropy in the human brain using double-diffusion encoding spectroscopy at 7T. Proc. Intl. Soc. Mag. Reson. Med. 27, 2019.

NALCIOGLU, O., CHO, Z. H., XIANG, Q. S. & AHN, C. B. 1986. Incoherent Flow Imaging. *Proc. SPIE 0671*.

NEEMAN, M., FREYER, J. P. & SILLERUD, L. O. 1991. A simple method for obtaining cross-term-free images for diffusion anisotropy studies in NMR microimaging. *Magn Reson Med,* 21**,** 138-43.

NERY, F., SZCZEPANKIEWICZ, F., KERKELA, L., HALL, M. G., KADEN, E., GORDON, I., THOMAS, D. L. & CLARK, C. A. 2019. In vivo demonstration of microscopic anisotropy in the human kidney using multidimensional diffusion MRI. *Magn Reson Med,* 82**,** 2160-2168.

NEWTON, I. 1701. A Scale of the Degrees of Heat (Translated from Latin). *Philosophical Transactions*.

NILSSON, M., ENGLUND, E., SZCZEPANKIEWICZ, F., VAN WESTEN, D. & SUNDGREN, P. C. 2018. Imaging brain tumour microstructure. *Neuroimage,* 182**,** 232-250.

NILSSON, M., LASIC, S., DROBNJAK, I., TOPGAARD, D. & WESTIN, C. F. 2017. Resolution limit of cylinder diameter estimation by diffusion MRI: The impact of gradient waveform and orientation dispersion. *NMR Biomed,* 30.

NILSSON, M., SZCZEPANKIEWICZ, F., BRABEC, J., TAYLOR, M., WESTIN, C. F., GOLBY, A., VAN WESTEN, D. & SUNDGREN, P. C. 2020a. Tensor-valued diffusion MRI in under 3 minutes: an initial survey of microscopic anisotropy and tissue heterogeneity in intracranial tumors. *Magn Reson Med,* 83**,** 608-620.

NILSSON, M., SZCZEPANKIEWICZ, F., VAN WESTEN, D. & HANSSON, O. 2015. Extrapolation-Based References Improve Motion and Eddy-Current Correction of High B-Value DWI Data: Application in Parkinson's Disease Dementia. *PLoS One,* 10**,** e0141825.

NILSSON, M., WESTIN, C.-F., BRABEC, J., LASIČ, S. & SZCZEPANKIEWICZ, F. A unified framework for analysis of time-dependent diffusion: numerical validation of a restriction-exchange correlation experiment. Proc. Intl. Soc. Mag. Reson. Med. 28, 2020b Virtual.

NING, L., NILSSON, M., LASIČ, S., WESTIN, C. F. & RATHI, Y. 2018. Cumulant expansions for measuring water exchange using diffusion MRI. *J Chem Phys,* 148**,** 074109.

NORHOJ JESPERSEN, S. 2018. White matter biomarkers from diffusion MRI. *J Magn Reson,* 291**,** 127-140.

NORRIS, D. G. & HUTCHISON, J. M. S. 1990. Concomitant magnetic field gradients and their effects on imaging at low magnetic field strengths. *Magnetic Resonance Imaging,* 8**,** 33-37.

NOVIKOV, D. S., FIEREMANS, E., JESPERSEN, S. N. & KISELEV, V. G. 2019. Quantifying brain microstructure with diffusion MRI: Theory and parameter estimation. *NMR Biomed,* 32**,** e3998.

NOVIKOV, D. S., JENSEN, J. H., HELPERN, J. A. & FIEREMANS, E. 2014. Revealing mesoscopic structural universality with diffusion. *Proc Natl Acad Sci U S A,* 111**,** 5088-93.

NOVIKOV, D. S., KISELEV, V. G. & JESPERSEN, S. N. 2018a. On modeling. *Magn Reson Med,* 79**,** 3172-3193.

NOVIKOV, D. S., REISERT, M. & KISELEV, V. G. 2018b. Effects of mesoscopic susceptibility and transverse relaxation on diffusion NMR. *J Magn Reson,* 293**,** 134-144.

NOVIKOV, D. S., VERAART, J., JELESCU, I. O. & FIEREMANS, E. 2018c. Rotationally-invariant mapping of scalar and orientational metrics of neuronal microstructure with diffusion MRI. *Neuroimage,* 174**,** 518-538.

NUNES, R. G., JEZZARD, P. & CLARE, S. 2005. Investigations on the efficiency of cardiac-gated methods for the acquisition of diffusion-weighted images. *J Magn Reson,* 177**,** 102-10.

OTT, M., BLAIMER, M., GRODZKI, D. M., BREUER, F. A., ROESCH, J., DORFLER, A., HEISMANN, B. & JAKOB, P. M. 2015. Acoustic-noise-optimized diffusion-weighted imaging. *MAGMA,* 28**,** 511-21.

OZAKI, M., INOUE, Y., MIYATI, T., HATA, H., MIZUKAMI, S., KOMI, S., MATSUNAGA, K. & WOODHAMS, R. 2013. Motion artifact reduction of diffusion-weighted MRI of the liver: use of velocity-compensated diffusion gradients combined with tetrahedral gradients. *J Magn Reson Imaging,* 37**,** 172-8.

OZARSLAN, E. & BASSER, P. J. 2007. MR diffusion - "diffraction" phenomenon in multi-pulse-field-gradient experiments. *J Magn Reson,* 188**,** 285-94.

PADHANI, A. R., LIU, G., MU-KOH, D., CHENEVERT, T. L., THOENY, H. C., TAKAHARA, T., DZIK-JURASZ, A., ROSS, B. D., VAN CAUTEREN, M., COLLINS, D., HAMMOUD, D. A., RUSTIN, G. J. S., TAOULI, B. & CHOYKE, P. L. 2009. Diffusion-Weighted Magnetic Resonance Imaging as a Cancer Biomarker: Consensus and Recommendations. *Neoplasia,* 11**,** 102-125.

PAMPEL, A., JOCHIMSEN, T. H. & MOLLER, H. E. 2010. BOLD background gradient contributions in diffusion-weighted fMRI--comparison of spin-echo and twice-refocused spin-echo sequences. *NMR Biomed,* 23**,** 610-8.

PARTRIDGE, S. C., NISSAN, N., RAHBAR, H., KITSCH, A. E. & SIGMUND, E. E. 2017. Diffusion-weighted breast MRI: Clinical applications and emerging techniques. *J Magn Reson Imaging,* 45**,** 337-355.

PIPE, J. 2010. Pulse Sequences for Diffusion-weighted MRI. *Diffusion MRI.* Oxford University Press.

PIPE, J. G. & CHENEVERT, T. L. 1991. A progressive gradient moment nulling design technique. *Magn Reson Med,* 19**,** 175-9.

POOT, D. H., DEN DEKKER, A. J., ACHTEN, E., VERHOYE, M. & SIJBERS, J. 2009. Optimal experimental design for Diffusion Kurtosis Imaging. *IEEE Trans. Med. Imaging,* 29**,** 819-829.

PORTNOY, S., FLINT, J. J., BLACKBAND, S. J. & STANISZ, G. J. 2013. Oscillating and pulsed gradient diffusion magnetic resonance microscopy over an extended b-value range: implications for the characterization of tissue microstructure. *Magn Reson Med,* 69**,** 1131-45.

PRICE, W. S. 1997. Pulsed-Field Gradient Nuclear Magnetic Resonance as a Tool for Studying Translational Diffusion: Part 1. Basic Theory. *Concepts in Magnetic Resonance,* 9**,** 299-336.

REESE, T. G., HEID, O., WEISSKOFF, R. M. & WEDEEN, V. J. 2003. Reduction of eddy-current-induced distortion in diffusion MRI using a twice-refocused spin echo. *Magn Reson Med,* 49**,** 177-82.

REILLY, J. P. 1989. Peripheral nerve stimulation by induced electric currents: exposure to time-varying magnetic fields. *Med Biol Eng Comput,* 27**,** 101-10.

REISERT, M., KISELEV, V. G. & DHITAL, B. 2019. A unique analytical solution of the white matter standard model using linear and planar encodings. *Magn Reson Med,* 81**,** 3819-3825.

REYMBAUT, A., MEZZANI, P., DE ALMEIDA MARTINS, J. P. & TOPGAARD, D. 2020. Accuracy and precision of statistical descriptors obtained from multidimensional diffusion signal inversion algorithms. *NMR Biomed***,** e4267.

REYNAUD, O. 2017. Time-Dependent Diffusion MRI in Cancer: Tissue Modeling and Applications. *Frontiers in Physics,* 5.

REYNAUD, O., WINTERS, K. V., HOANG, D. M., WADGHIRI, Y. Z., NOVIKOV, D. S. & KIM, S. G. 2016. Pulsed and oscillating gradient MRI for assessment of cell size and extracellular space (POMACE) in mouse gliomas. *NMR Biomed*.

ROSCH, J., OTT, M., HEISMANN, B., DOERFLER, A., ENGELHORN, T., SEMBRITZKI, K. & GRODZKI, D. M. 2016. Quiet diffusion-weighted head scanning: Initial clinical evaluation in ischemic stroke patients at 1.5T. *J Magn Reson Imaging,* 44**,** 1238-1243.

RUDRAPATNA, U., PARKER, G. D., ROBERTS, J. & JONES, D. K. 2020. A comparative study of gradient nonlinearity correction strategies for processing diffusion data obtained with ultra-strong gradient MRI scanners. *Magn Reson Med*.

SCHMITT, F. The Gradient System. Understanding Gradients from an EM Perspective: (Gradient Linearity, Eddy Currents, Maxwell Terms & Peripheral Nerve Stimulation). Proc. Intl. Soc. Magn. Reson. Med. 21, 2013 Salt Lake City, UT, USA.

SCHULTE, R. F. & NOESKE, R. 2015. Peripheral nerve stimulation-optimal gradient waveform design. *Magn Reson Med,* 74**,** 518-22.

SEN, P. N. 2004. Time-dependent diffusion coefficient as a probe of geometry. *Concepts in Magnetic Resonance,* 23A**,** 1-21.

SETSOMPOP, K., KIMMLINGEN, R., EBERLEIN, E., WITZEL, T., COHEN-ADAD, J., MCNAB, J. A., KEIL, B., TISDALL, M. D., HOECHT, P., DIETZ, P., CAULEY, S. F., TOUNTCHEVA, V., MATSCHL, V., LENZ, V. H., HEBERLEIN, K., POTTHAST, A., THEIN, H., VAN HORN, J., TOGA, A., SCHMITT, F., LEHNE, D., ROSEN, B. R., WEDEEN, V. & WALD, L. L. 2013. Pushing the limits of in vivo diffusion MRI for the Human Connectome Project. *Neuroimage,* 80**,** 220-33.

SHEMESH, N., JESPERSEN, S. N., ALEXANDER, D. C., COHEN, Y., DROBNJAK, I., DYRBY, T. B., FINSTERBUSCH, J., KOCH, M. A., KUDER, T., LAUN, F., LAWRENZ, M., LUNDELL, H., MITRA, P. P., NILSSON, M., OZARSLAN, E., TOPGAARD, D. & WESTIN, C. F. 2016. Conventions and nomenclature for double diffusion encoding NMR and MRI. *Magn Reson Med,* 75**,** 82-7.

SHEMESH, N., OZARSLAN, E., ADIRI, T., BASSER, P. J. & COHEN, Y. 2010a. Noninvasive bipolar double-pulsed-field-gradient NMR reveals signatures for pore size and shape in polydisperse, randomly oriented, inhomogeneous porous media. *J Chem Phys,* 133**,** 044705.

SHEMESH, N., OZARSLAN, E., BASSER, P. J. & COHEN, Y. 2009. Measuring small compartmental dimensions with low-q angular double-PGSE NMR: The effect of experimental parameters on signal decay. *J Magn Reson,* 198**,** 15-23.

SHEMESH, N., ÖZARSLAN, E., KOMLOSH, M. E., BASSER, P. J. & COHEN, Y. 2010b. From single-pulsed field gradient to double-pulsed field gradient MR: gleaning new microstructural information and developing new forms of contrast in MRI. *NMR Biomed,* 23**,** 757-80.

SJÖLUND, J., SZCZEPANKIEWICZ, F., NILSSON, M., TOPGAARD, D., WESTIN, C. F. & KNUTSSON, H. 2015. Constrained optimization of gradient waveforms for generalized diffusion encoding. *J Magn Reson,* 261**,** 157-168.

SKARE, S. & ANDERSSON, J. L. R. 2001. On the effects of gating in diffusion imaging of the brain using single shot EPI. *Magnetic Resonance Imaging,* 19**,** 1125–1128.

SMINK, J., PLATTEL, G.-J., HARVEY, P. R. & LIMPENS, P. General Method for Acoustic Noise Reduction by Avoiding Resonance Peaks. Proc. Int. Soc. Magn. Reson. Med. 15, 2007 Berlin, Germany.

STANISZ, G. J., SZAFER, A., WRIGHT, G. A. & HENKELMAN, R. M. 1997. An analytical model of restricted diffusion in bovine optic nerve. *Magn Reson Med,* 37**,** 103-11.

STEJSKAL, E. O. 1965. Use of Spin Echoes in a Pulsed Magnetic-Field Gradient to Study Anisotropic, Restricted Diffusion and Flow. *The Journal of Chemical Physics,* 43**,** 3597.

STEJSKAL, E. O. & TANNER, J. E. 1965. Spin Diffusion Measurement: Spin echoes in the Presence of a Time-Dependent Field Gradient. *the journal of chemical physics,* 42**,** 288-292.

STEPISNIK, J. 1993. Time-Dependent Self-Diffusion by Nmr Spin-Echo. *Physica B,* 183**,** 343-350.

STEPISNIK, J. 1999. Validity limits of Gaussian approximation in cumulant expansion for diffusion attenuation of spin echo. *Physica B,* 270**,** 110-117.

SUNDGREN, P. C., DONG, Q., GOMEZ-HASSAN, D., MUKHERJI, S. K., MALY, P. & WELSH, R. 2004. Diffusion tensor imaging of the brain: review of clinical applications. *Neuroradiology,* 46**,** 339-50.

SZCZEPANKIEWICZ, F. 2016. *Imaging diffusional variance by MRI: The role of tensor-valued diffusion encoding and tissue heterogeneity.* Ph.D. Thesis, Lund University.

SZCZEPANKIEWICZ, F., EICHNER, C., ANWANDER, A., WESTIN, C.-F. & PAQUETTE, M. The impact of gradient non-linearity on Maxwell compensation when using asymmetric gradient waveforms for tensor-valued diffusion encoding. Proc. Intl. Soc. Mag. Reson. Med. 28, 2020a Virtual.

SZCZEPANKIEWICZ, F., HOGE, S. & WESTIN, C. F. 2019a. Linear, planar and spherical tensor-valued diffusion MRI data by free waveform encoding in healthy brain, water, oil and liquid crystals. *Data Brief,* 25**,** 104208.

SZCZEPANKIEWICZ, F., LASIC, S., NILSSON, C., LUNDELL, H., WESTIN, C. F. & TOPGAARD, D. Is spherical diffusion encoding rotation invariant? An investigation of diffusion timedependence in the healthy brain. Proc. Intl. Soc. Mag. Reson. Med. 27, 2019b.

SZCZEPANKIEWICZ, F., LASIČ, S., VAN WESTEN, D., SUNDGREN, P. C., ENGLUND, E., WESTIN, C. F., STÅHLBERG, F., LÄTT, J., TOPGAARD, D. & NILSSON, M. 2015. Quantification of microscopic diffusion anisotropy disentangles effects of orientation dispersion from microstructure: Applications in healthy volunteers and in brain tumors. *Neuroimage,* 104**,** 241-252.

SZCZEPANKIEWICZ, F., SJÖLUND, J., DALL'ARMELLINA, E., PLEIN, S., SCHNEIDER, J. E., TEH, I. & WESTIN, C.-F. 2020b. Motion-compensated gradient waveforms for tensor-valued diffusion encoding by constrained numerical optimization. *Magnetic Resonance in Medicine,* (accepted, in review).

SZCZEPANKIEWICZ, F., SJÖLUND, J., STÅHLBERG, F., LÄTT, J. & NILSSON, M. 2019c. Tensor-valued diffusion encoding for diffusional variance decomposition (DIVIDE): Technical feasibility in clinical MRI systems. *PLoS One,* 14**,** e0214238.

SZCZEPANKIEWICZ, F., VAN WESTEN, D., ENGLUND, E., WESTIN, C. F., STAHLBERG, F., LATT, J., SUNDGREN, P. C. & NILSSON, M. 2016. The link between diffusion MRI and tumor heterogeneity: Mapping cell eccentricity and density by diffusional variance decomposition (DIVIDE). *Neuroimage,* 142**,** 522-532.

SZCZEPANKIEWICZ, F., WESTIN, C. F. & NILSSON, M. 2019d. Maxwell-compensated design of asymmetric gradient waveforms for tensor-valued diffusion encoding. *Magn Reson Med*.

TAN, E. T., HUA, Y., FIVELAND, E. W., VERMILYEA, M. E., PIEL, J. E., PARK, K. J., HO, V. B. & FOO, T. K. F. 2020. Peripheral nerve stimulation limits of a high amplitude and slew rate magnetic field gradient coil for neuroimaging. *Magn Reson Med,* 83**,** 352-366.

TAN, E. T., MARINELLI, L., SLAVENS, Z. W., KING, K. F. & HARDY, C. J. 2013. Improved correction for gradient nonlinearity effects in diffusion-weighted imaging. *J Magn Reson Imaging,* 38**,** 448-53.

TAOULI, B., BEER, A. J., CHENEVERT, T., COLLINS, D., LEHMAN, C., MATOS, C., PADHANI, A. R., ROSENKRANTZ, A. B., SHUKLA-DAVE, A., SIGMUND, E., TANENBAUM, L., THOENY, H., THOMASSIN-NAGGARA, I., BARBIERI, S., CORCUERA-SOLANO, I., ORTON, M., PARTRIDGE, S. C. & KOH, D. M. 2016. Diffusion-weighted imaging outside the brain: Consensus statement from an ISMRM-sponsored workshop. *J Magn Reson Imaging,* 44**,** 521-40.

TAX, C. M. W., SZCZEPANKIEWICZ, F., NILSSON, M. & JONES, D. K. 2020. The dot-compartment revealed? Diffusion MRI with ultra-strong gradients and spherical tensor encoding in the living human brain. *Neuroimage***,** 116534.

THOMAS, C. & BAKER, C. I. 2013. Teaching an adult brain new tricks: a critical review of evidence for training-dependent structural plasticity in humans. *Neuroimage,* 73**,** 225-36.

TOPGAARD, D. 2013. Isotropic diffusion weighting in PGSE NMR: Numerical optimization of the q-MAS PGSE sequence. *Microporous and Mesoporous Materials,* 178**,** 60-63.

TOPGAARD, D. 2016. Director orientations in lyotropic liquid crystals: diffusion MRI mapping of the Saupe order tensor. *Phys Chem Chem Phys,* 18**,** 8545-53.

TOPGAARD, D. 2017. Multidimensional diffusion MRI. *J Magn Reson,* 275**,** 98-113.

TORREY, H. C. 1956. Bloch Equations with Diffusion Terms. *Physical Review,* 104**,** 563-565.

TOURNIER, J. D. 2019. Diffusion MRI in the brain – Theory and concepts. *Progress in Nuclear Magnetic Resonance Spectroscopy,* 112-113**,** 1-16.

TRUFFET, R., RAFAEL-PATINO, J., GIRARD, G., PIZZOLATO, M., BARILLOT, C., THIRAN, J.-P. & CARUYER, E. An evolutionary framework for microstructure-sensitive generalized diffusion gradient waveforms. International Conference on Medical Image Computing and Computer Assisted Intervention, 2020 Lima, Peru.

TSIEN, C., CAO, Y. & CHENEVERT, T. 2014. Clinical applications for diffusion magnetic resonance imaging in radiotherapy. *Semin Radiat Oncol,* 24**,** 218-26.

TURNER, R. 1993. Gradient coil design: a review of methods. *Magn Reson Imaging,* 11**,** 903-20.

VALETTE, J., GIRAUDEAU, C., MARCHADOUR, C., DJEMAI, B., GEFFROY, F., GHALY, M. A., LE BIHAN, D., HANTRAYE, P., LEBON, V. & LETHIMONNIER, F. 2012. A new sequence for single-shot diffusion-weighted NMR spectroscopy by the trace of the diffusion tensor. *Magn Reson Med,* 68**,** 1705-12.

VAN VAALS, J. J. & BERGMAN, A. H. 1990. Optimization of Eddy-Current Compensatio. *Journal of Magnetic Resonance,* 90**,** 52-70.

WEDEEN, V. J., DAI, G., TSENG, W.-Y. I., WANG, R. & BENNER, T. Diffusion encoding with 2D gradient trajectories yields natural contrast for 3D fiber orientation. Proc. Intl. Soc. Mag. Reson. Med. 14, 2006.

WEDEEN, V. J., ROSENE, D. L., WANG, R., DAI, G., MORTAZAVI, F., HAGMANN, P., KAAS, J. H. & TSENG, W. Y. 2012. The geometric structure of the brain fiber pathways. *Science,* 335**,** 1628-34.

WEIDLICH, D., ZAMSKIY, M., MAEDER, M., RUSCHKE, S., MARBURG, S. & KARAMPINOS, D. C. 2020. Reduction of vibration-induced signal loss by matching mechanical vibrational states: Application in high b-value diffusion-weighted MRS. *Magn Reson Med,* 84**,** 39-51.

WEIGER, M., OVERWEG, J., ROSLER, M. B., FROIDEVAUX, R., HENNEL, F., WILM, B. J., PENN, A., STURZENEGGER, U., SCHUTH, W., MATHLENER, M., BORGO, M., BORNERT, P., LEUSSLER, C., LUECHINGER, R., DIETRICH, B. E., REBER, J., BRUNNER, D. O., SCHMID, T., VIONNET, L. & PRUESSMANN, K. P. 2018. A high-performance gradient insert for rapid and short-T2 imaging at full duty cycle. *Magn Reson Med,* 79**,** 3256-3266.

WESTIN, C. F., KNUTSSON, H., PASTERNAK, O., SZCZEPANKIEWICZ, F., ÖZARSLAN, E., VAN WESTEN, D., MATTISSON, C., BOGREN, M., O'DONNELL, L. J., KUBICKI, M.,

TOPGAARD, D. & NILSSON, M. 2016. Q-space trajectory imaging for multidimensional diffusion MRI of the human brain. *Neuroimage,* 135**,** 345-62.

WESTIN, C. F., MAIER, S. E., MAMATA, H., NABAVI, A., JOLESZ, F. A. & KIKINIS, R. 2002. Processing and visualization for diffusion tensor MRI. *Medical Image Analysis,* 6**,** 93-108.

WESTIN, C. F., SZCZEPANKIEWICZ, F., PASTERNAK, O., ÖZARSLAN, E., TOPGAARD, D., KNUTSSON, H. & NILSSON, M. 2014. Measurement tensors in diffusion MRI: Generalizing the concept of diffusion encoding. *Med Image Comput Comput Assist Interv,* 17 (Pt 5)**,** 217-225.

WINTHER ANDERSEN, K., LASIČ, S., LUNDELL, H., NILSSON, M., TOPGAARD, D., SELLEBJERG, F., SZCZEPANKIEWICZ, F., ROMAN SIEBNER, H., BLINKENBERG, M. & DYRBY, T. 2020. Disentangling white-matter damage from physiological fiber orientation dispersion in multiple sclerosis. . *Brain Communications (in press).*

WONG, E. C., COX, R. W. & SONG, A. W. 1995. Optimized isotropic diffusion weighting. *Magn Reson Med,* 34**,** 139-43.

VOS, S. B., TAX, C. M., LUIJTEN, P. R., OURSELIN, S., LEEMANS, A. & FROELING, M. 2017. The importance of correcting for signal drift in diffusion MRI. *Magn Reson Med,* 77**,** 285-299.

YANG, G. & MCNAB, J. A. 2018. Eddy current nulled constrained optimization of isotropic diffusion encoding gradient waveforms. *Magn Reson Med*.

ZATORRE, R. J., FIELDS, R. D. & JOHANSEN-BERG, H. 2012. Plasticity in gray and white: neuroimaging changes in brain structure during learning. *Nat Neurosci,* 15**,** 528-36.

ZHENG, G. & PRICE, W. S. 2007. Suppression of background gradients in (B0 gradient-based) NMR diffusion experiments. *Concepts in Magnetic Resonance Part A,* 30A**,** 261-277.

ZHOU, X. J., TAN, S. G. & BERNSTEIN, M. A. 1998. Artifacts induced by concomitant magnetic field in fast spin-echo imaging. *Magn Reson Med,* 40**,** 582-91.

ÖZARSLAN, E. 2009. Compartment shape anisotropy (CSA) revealed by double pulsed field gradient MR. *J Magn Reson,* 199**,** 56-67.

ÖZARSLAN, E. & BASSER, P. J. 2008. Microscopic anisotropy revealed by NMR double pulsed field gradient experiments with arbitrary timing parameters. *J Chem Phys,* 128**,** 154511.